\documentclass[lettersize,journal]{IEEEtran}
\usepackage{amsmath,amsfonts}
\usepackage{algorithmic}
\usepackage{algorithm}
\usepackage{array}
\usepackage[caption=false,font=normalsize,labelfont=sf,textfont=sf]{subfig}
\usepackage{textcomp}
\usepackage{stfloats}
\usepackage{url}
\usepackage{verbatim}
\usepackage{graphicx}
\usepackage{cite}
\usepackage{multirow}
\usepackage{tabularx}
\usepackage{makecell}
\usepackage{booktabs}
\usepackage{enumitem}
\usepackage{hyperref} 
\newcommand{\tabincell}[2]{\begin{tabular}{@{}#1@{}}#2\end{tabular}}
\hypersetup
{
 unicode=false,     
 pdftoolbar=true,    
 pdfmenubar=true,    
 pdffitwindow=false,   
 pdfstartview={FitH},  
 pdftitle={My title},  
 pdfauthor={Author},   
 pdfsubject={Subject},  
 pdfcreator={Creator},  
 pdfproducer={Producer}, 
 pdfkeywords={keywords}, 
 pdfnewwindow=true,   
 colorlinks=true,    
 linkcolor=red,     
 citecolor=blue,    
 filecolor=magenta,   
 urlcolor=cyan      
}
\usepackage[switch]{lineno}
\begin{document}

\title{Game Theoretic Application to Intersection Management: A Literature Review}

\author{Ziye~Qin,~\IEEEmembership{Student Member,~IEEE,}
        Ang~Ji,~\IEEEmembership{Member,~IEEE,}
        Zhanbo~Sun,
        Guoyuan~Wu,~\IEEEmembership{Senior~Member,~IEEE,}
        Peng~Hao,~\IEEEmembership{Member,~IEEE,}
        and~Xishun~Liao,~\IEEEmembership{Student Member,~IEEE}

\thanks{Ziye Qin, Ang Ji, and Zhanbo Sun (Corresponding author) are with the School of Transportation and Logistics, Southwest Jiaotong University, Chengdu, Sichuan, 611756, China, and the National Engineering Laboratory of Integrated Transportation Big Data Application Technology, Southwest Jiaotong University, Chengdu, Sichuan, 611756, China. (Email: ziye.qin@my.swjtu.edu.cn,  ang.ji@swjtu.edu.cn, zhanbo.sun@home.swjtu.edu.cn; Tel: 86-28-87600156; Fax: 86-28-87600156).}
\thanks{Ziye Qin,  Guoyuan Wu, and Peng Hao are with the College of Engineering, Center for Environmental Research and Technology, University of California at Riverside, Riverside, CA 92507 USA. (Email: gywu@cert.ucr.edu, haop@cert.ucr.edu).}

\thanks{Xishun Liao is with the University of California at Los Angeles, Los Angeles, CA 90095 USA. (Email: liaoxishun@gmail.com).}

}



\maketitle

\begin{abstract}
    The emergence of vehicle-to-everything (V2X) technology offers new insights into intersection management. This, however, has also presented new challenges, such as the need to understand and model the interactions of traffic participants, including their competition and cooperation behaviors. Game theory has been widely adopted to study rationally selfish or cooperative behaviors during interactions and has been applied to advanced intersection management. In this paper, we review the application of game theory to intersection management and sort out relevant studies under various levels of intelligence and connectivity. First, the problem of urban intersection management and its challenges are briefly introduced. The basic elements of game theory specifically for intersection applications are then summarized. Next, we present the game-theoretic models and solutions that have been applied to intersection management. Finally, the limitations and potential opportunities for subsequent studies within the game-theoretic application to intersection management are discussed.
\end{abstract}


\section{Introduction}
\IEEEPARstart{I}{ntersections} within urban traffic networks are subject to traffic conflicts, leading to prolonged vehicle delay and elevated energy consumption \cite{mirheli2019consensus,yu2018integrated,qian2019toward}. Innovative technologies such as human-machine interface (HMI), and vehicle-to-everything (V2X) communications facilitate the collection and sharing of vehicular and trip information, such as origin and destination, vehicle trajectories, and personal attributes \cite{dey2016vehicle,sun2020cooperative,chen2022milestones}. These advancements unlock a significant number of opportunities for more efficient intersection management. For example, traffic signal controllers can utilize more comprehensive vehicle information to optimize signal phase and timing, and minimize overall traffic delays. Since the introduction of the autonomous intersection management (AIM) system by Dresner and Stone, there have been increasing efforts to explore the applications of emerging technologies in this area \cite{dresner2008multiagent}. Simultaneously, management strategies have shifted from focusing only on macroscopic measurements (e.g., traffic efficiency) to microscopic considerations such as experiences and intentions of traffic participants \cite{wang2022social}.

In the context of management and control, intersections can be classified as signalized and un-signalized intersections \cite{pruekprasert2019decision}. The optimization of signalized intersections typically involves the allocation of the signal phase and timing (SPaT). Through the optimization of SPaT, bus rapid transit (BRT) priority strategies can be effectively executed \cite{liu2020reservation}.  With the advancement of cooperative adaptive cruise control (CACC) technology, a compatible SPaT plan has been demonstrated to improve intersection capacity \cite{liu2019traffic}. In the partially/fully connected automated vehicle (CAV) environment, signal optimization can be coupled with vehicle trajectory planning \cite{guo2019joint,feng2018spatiotemporal}. The co-optimization of SPaT information and vehicle speed has the potential to concurrently enhance eco-driving practices and traffic efficiency \cite{hao2018eco,sun2022eco}. Even given SPAT information, optimizing vehicle trajectories can balance traffic efficiency and fuel economy \cite{lin2017autonomous,yang2020eco}. Meanwhile, the management of signalized intersections, considering the stochastic behavior inherent to human-operated vehicles (HVs), is attracting heightened interest \cite{xiong2021speed,chen2021mixed}. In an ideal environment characterized by a full penetration of CAVs, the utilization of traffic signals may become redundant (i.e., un-signalized intersections). The main issues in managing un-signalized intersections pertain to right-of-way allocation and collision-free vehicle trajectory planning \cite{li2020game,hu2021constraint,chen2022conflict,luo2023real}. The trajectory planning for cyclists, a vulnerable demographic at un-signalized intersections, is garnering attention. This concern arises from the conflicts that cyclists may encounter both within their group and with motorized vehicles \cite{huang2009cyclists}. For all types of interactions, the optimization cannot be solely considered from the perspective of traffic efficiency; factors such as the value of time, time-to-collision (TTC), and fuel consumption are also frequently investigated  \cite{isukapati2017accommodating,niroumand2020joint,lin2021pay}. Several review articles have summarized the existing methodologies on intersection management and proposed promising future directions. For example, Zhao et al. discussed the application of computational intelligence in traffic signal control that can address the large complex nonlinear problems in urban transportation systems \cite{zhao2011computational}. Araghi et al. also reviewed computational intelligence methods for traffic signal control and compared the performance of various learning methods \cite{araghi2015review}. Wang et al. surveyed the history of traffic self-adaptive control systems, pointed out the deficiencies, and expressed the expectations for multi-agent reinforcement learning in this field \cite{wang2018review}. Yau et al. surveyed the application of reinforcement learning algorithms to traffic signal control \cite{yau2017survey}. Florin and Olariu reviewed three adaptive traffic signal control strategies for various levels of vehicular communications, including: (i) no vehicle communication; (ii) wireless transmission of vehicle status data; and (iii) on-board computation to optimize traffic signals \cite{florin2015survey}. Chen and Englund surveyed cooperative intersection management strategies under both signalized and un-signalized intersections and summarized related research projects worldwide \cite{chen2015cooperative}. Studies related to CAV-enabled intelligent intersection management were summarized in \cite{rios2016survey,guo2019urban,khayatian2020survey}. Namazi et al. provided a systematic review of intelligent intersection management in mixed traffic including autonomous vehicle (AV) and human-operated vehicle (HV), also with high expectations for artificial intelligence (AI)-enabled traffic management in the future \cite{namazi2019intelligent}. Al-Turki et al. concluded signalized intersection control methods in mixed traffic and proposed an alternative machine learning (ML)-based solution to the design of intelligent adaptive traffic signal controllers \cite{al2022signalized}. Additionally, the existing challenges and potential future directions were outlined. Shirazi and Morris reviewed the behavior at intersections and the safety analysis for vehicles, drivers, and pedestrians, and made suggestions for the safety management of intersections \cite{shirazi2016looking}. Zhong et al. provided a summary of AIM research in terms of the corridor coordination layer, the intersection management layer, and the vehicle control layer \cite{zhong2020autonomous}.  Iliopoulou et al. summarized the market-inspired allocation of spatial-temporal resources at intersections within the context of the connected vehicle (CV) environment \cite{iliopoulou2022survey}. The aforementioned reviews highlighted the significant accomplishments in the development of technical tools for intersection management. Nevertheless, most traffic management strategies, when scrutinized from an optimization standpoint, have been developed in an altruistic manner, that is, any player might have to sacrifice his or her own interest to achieve the global optimum. For example, vehicles are often obliged to endure waiting times for the larger collective benefits, without any compensation. Therefore, enhancing our understanding of the needs of traffic participants, the intricacies of their interactive behaviors, and the delicate equilibrium between individual gains and collective benefits are worthy subjects for continued exploration. Toward this end, a comprehensive summary of interaction analysis between decision-makers at intersections holds promise in facilitating the research of more user-friendly management strategies.

Consequently, there is an urgent demand to identify a suitable tool for analyzing individual preferences and interaction behaviors to guarantee fair decision-making in future intersection management. Game theory – known as a classic mathematical method to model conflicts and cooperation among decision-makers, is commonly used to study the behaviors of traffic participants \cite{myerson1997game}. A game consists of three basic elements: players, strategies for players, and payoff functions that describe how much payoff or utility each player can obtain under various environment states and joint actions chosen by the players \cite{ji2020review,kamhoua2021game}. Game theory has demonstrated its potential for managing traffic in various scenarios such as traffic congestion management \cite{ahmad2023game}, lane-changing \cite{zhang2019game,ji2020estimating,ji2023pricing,wang2022modeling,karimi2023level}, ramp-merging \cite{Sun2022micro,liao2021game,li2023simulation,sun2020microscopic}, and intersection management \cite{sayin2018information,wang2019enabling}. In lane-changing and ramp-merging games, the number of participants may be much less than that at a standard four-arm intersection where the interactions of traffic participants are more complex. In terms of the operational strategies, discrete actions such as acceleration/deceleration during lane-changing and yield/not yield during ramp-merging may be sufficient in decision-making modeling. However, as the diversity and number of participants grow at intersections, lateral and longitudinal joint strategies are needed for adapting to complex situations. An example of this complexity is the coordination of the green phase of traffic signals. Factors such as TTC and travel time are widely used to construct the payoff function from the perspective of players' safety and driving efficiency \cite{wang2021game}. In scenarios where players are traffic operators, traffic signal delays and queue lengths are frequently considered \cite{lloret2016envy,abdelghaffar2019development}. In addition, heterogeneous trip characteristics, such as trip purpose and distance, result in diversified preferences of time, safety, and other measures \cite{Sun2022micro, sun2021modeling}. For example, catching a flight generally has a higher value of time compared to shopping. Consequently, the payoff functions should be more specific. Another concern is that the continuous flow approaching the intersection makes the problem of passing sequence optimization even more complex, which can be NP-hard \cite{zhang2014analysis,zhang2021trajectory}. Game theory serves as a valuable instrument for decomposing complex interactions, modeling decision-making processes, analyzing the behavior of traffic participants, and helping take reasonable actions to manage the aforementioned issues. Specifically, game theory recognizes and accommodates user heterogeneity in the first category of issues, allowing for tailored personalization of payoff function compositions in line with individual characteristics. For the second category, a synergistic combination of game theory and multi-agent reinforcement learning (MARL) offers a potent solution, effectively resolving the intricate challenges presented.

Great efforts have been made to apply game-theoretic models for intersection management. The diversity of models makes it challenging to outline the existing studies, so a clear lineage of game theoretic approaches for intersection management needs to be identified. Additionally, most studies simplified intersection games by assuming a complete information environment. In reality, self-interested players may exploit misreported private information to increase their payoff, making it more practical to address the problem in an incomplete information environment. In a nutshell, despite the well-established conventional game models for intersection management, there are several challenges to be tackled in practical applications, such as how to find an appropriate game model and how to incentivize traffic participants to take part in the games. 

Based on the author's understanding, there have been few studies that provide a comprehensive summary of recent developments in game theory applications to intersections, especially on the behavior analysis and the modeling of traffic participants to enhance intersection management. Moreover, as connected and automated technologies gradually integrate into transportation systems, the emphasis on individual travel needs grows correspondingly. In this context, there is an evident demand for methods capable of balancing the requirements of traffic participants, guaranteeing their respective interests. Game theory provides a fitting solution to this requirement. Employing game-theoretic methods to manage intersection areas with multiple conflict points holds significant promise. To fill research gaps and facilitate knowledge and development for future relevant research, the paper provides a review on the decision-making problems of intersection management and aims to investigate the applications of game theory to intersections. The main contributions of this paper include: (i) summarizing the existing applications of game theory for intersections, including the intersection circumstances for deploying game theory, the basic elements of game theory, and the various types of game theory models and solutions; and (ii) highlighting existing issues and future directions in combining game theory and intersection management.

\section{Applications of Game Theory to Intersection Management}
\begin{figure*}[!t]
  \centering
  \includegraphics[width=\textwidth]{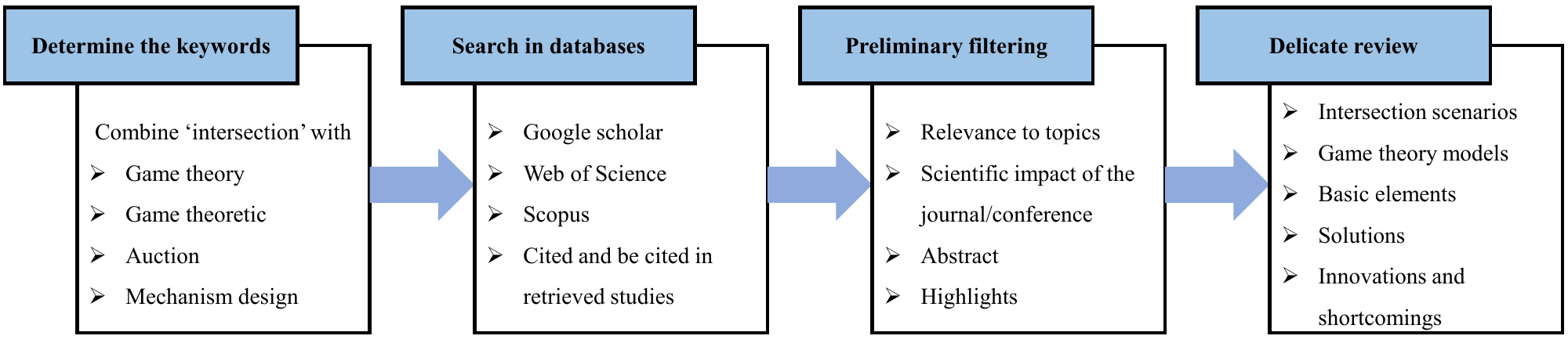}
  \caption{Literature retrieval and analysis workflow for collecting relevant papers}
  \label{fig1}
\end{figure*}

To collect relevant literature on game-theoretic applications to intersections, we presented a workflow for systematic literature retrieval and analysis, illustrated in Fig. \ref{fig1}. This methodology is versatile and can be adapted to trace advancements in diverse research fields. We began by identifying critical keywords associated with our research focus, incorporating terms such as ``game theory", ``game theoretic", ``auction", and ``mechanism design", combined with ``intersection", to retrieve relevant studies. This was followed by a keyword search in prominent scientific repositories like Google Scholar, Web of Science, and Scopus. We also followed up on articles citing the selected publications. Subsequently, we undertook a preliminary screening step to offset potential inaccuracies stemming from keyword searches. This involved a brief review of the topics, abstracts, highlights, and the prestige of the journals in which these studies were disseminated. Finally, we provide an in-depth review and summary of the game theory framework as applied to intersection management. This analysis is methodically presented, encompassing the types of intersections examined, the game theoretic modeling procedures, and both the innovations and limitations observed. An extensive search using specific keywords in databases yielded a significant number of pertinent studies. However, initial scrutiny and subsequent in-depth reviews identified instances of duplicative content. Finally, guided by the criteria presented in Fig. \ref{fig1}, we selected 88 articles, published between 2005 and 2023, for inclusion in this review. 

\subsection{An Overview of Intersection Management}
\begin{figure*}[!t]
  \centering
  \includegraphics[width=0.9\textwidth]{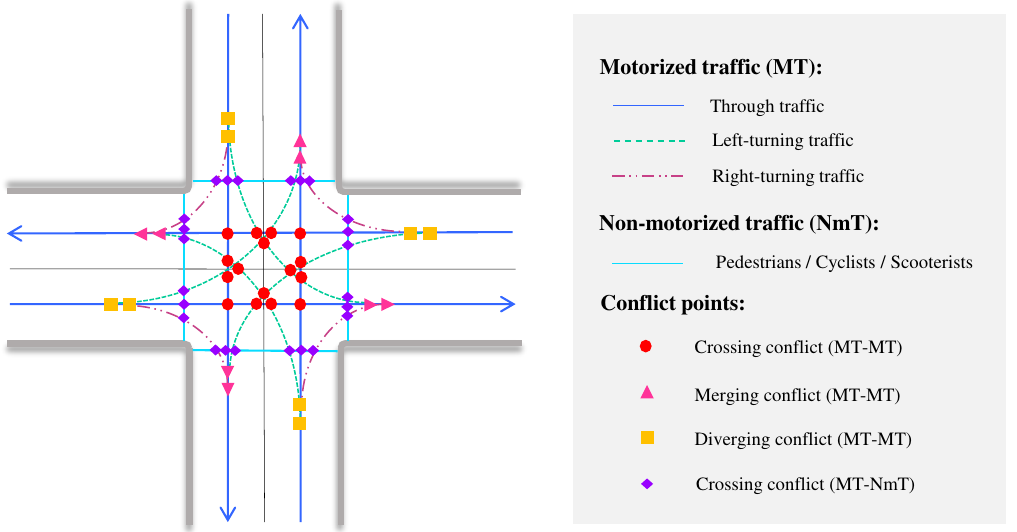}
  \caption{Illustration of conflict points for a standard four-leg intersection}
  \label{fig2}
\end{figure*}

Intersections serve as the critical points in urban transportation networks where multiple roadways converge horizontally; they also represent concentrated conflict areas, resulting in a higher incidence of collisions compared to the other parts of urban transportation networks \cite{wolshon2016traffic}. Fig. \ref{fig2} presents a standard four-leg intersection with three movements along each approach, including through traffic, left-turning traffic, and right-turning traffic. When traffic flows from different directions pass through the same section of the intersection, there are some potential collision areas (defined as ``conflict points") \cite{zhu2015linear}. Conflict points at intersections can be categorized as merging conflict points, diverging conflict points, and crossing conflict points \cite{hancock2013policy}. Merging conflict points arise when two paths converge or come together into a single path. Conversely, diverging conflict points emerge when one path splits into two or more paths. Lastly, crossing conflict points are observed when two paths intersect each other perpendicularly or at some angle, resulting in the potential for one road user to cross directly into the path of another. There exist thirty-two (32) conflict points among motorized traffic (MT) in the given scenario. This features eight (8) merging and eight (8) diverging conflict points, where rear-ended and sideswipe collisions typically occur. Additionally, there are sixteen (16) crossing conflict points inside the intersection, with twelve (12) of them associated with left-turn movements. The remaining four (4) crossing conflict points involve the through movements on two adjacent approaches, where angle collisions may occur \cite{rodegerdts2004signalized}. Non-motorized traffic (NmT), as the vulnerable group, encounters twenty-four (24) conflict points interacting with motorized traffic. Conventional intersection management relies on the separation of time-space conflicts among approaching vehicles, using traffic signal control and lane channelization, which also limits intersection capacity. To balance efficiency and safety, many studies have explored game-theoretic approaches for intersection management, which have proven to be a valuable tool for characterizing such interactions, particularly in the decision-making of CAVs \cite{Bui2017, Zhao2019, Khanjary2013}. 

\begin{figure*}[!t]
  \centering
  \includegraphics[width=6.5in]{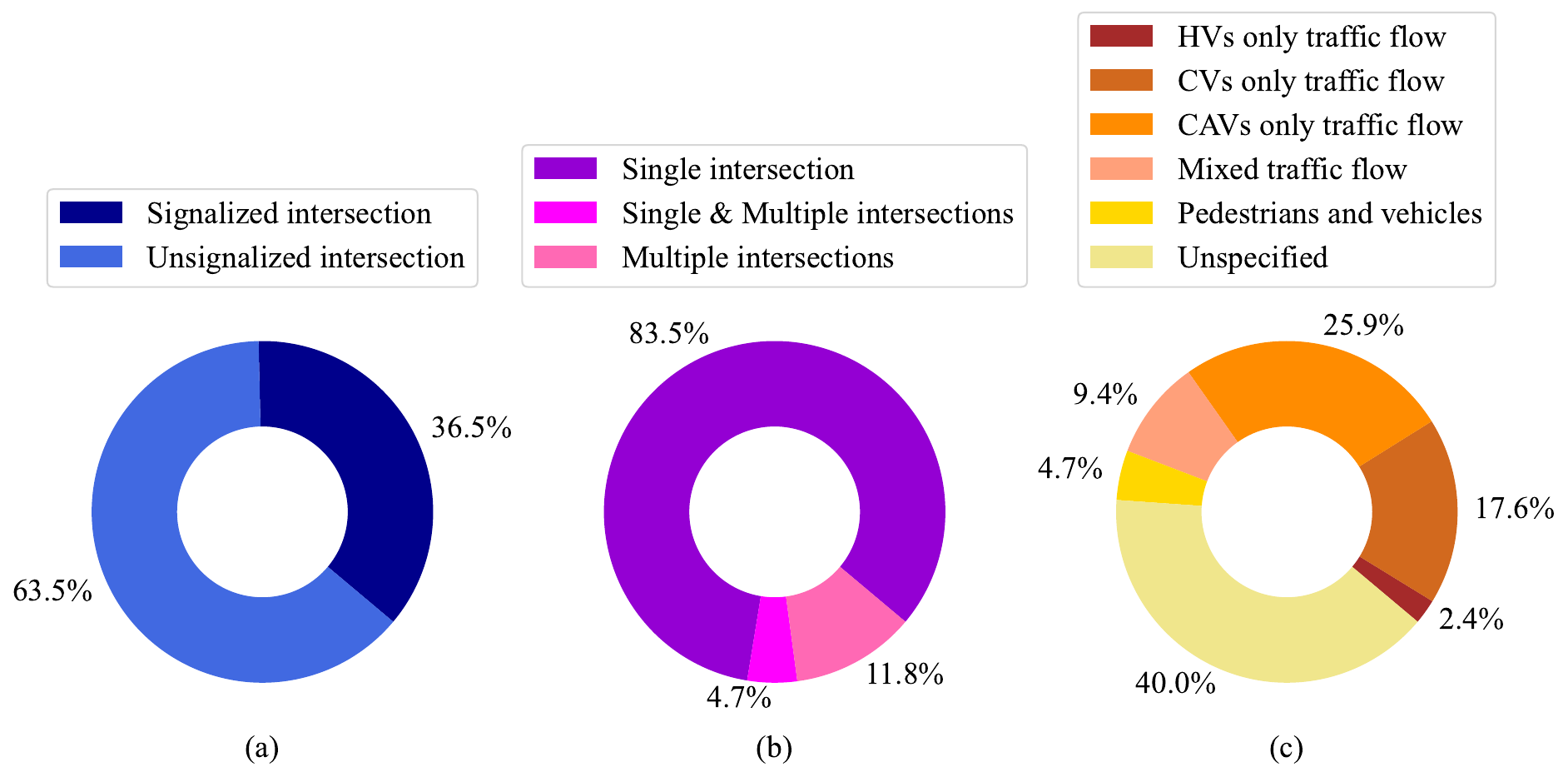}
  \caption{Applications of game theory under various intersection circumstances}
  \label{fig3}
\end{figure*}

To better understand the operation mechanism of intersections and the interactions among traffic participants, we revisited the intersection management solutions with the application of game theory. The intersection circumstances in game-theoretic studies can be classified based on several criteria: (i) the presence or absence of traffic signals: signalized and un-signalized intersections; (ii) the number of intersections under consideration: single, isolated intersections and/or multiple intersections; (iii) the type of traffic composition: traffic flow with HVs only, CVs only, CAVs only, mixed traffic with pedestrians and vehicles. A brief overview of the studies conducted in the various circumstances is presented in Fig. \ref{fig3}.  More than half of the game theory-related studies (63.5\%) have focused on the interactive behaviors among vehicles at un-signalized intersections. The multi-intersection coordination problem accounts for only 16.5\% of studies, which still needs further attention. A major chunk of studies focuses on the traffic flow with CVs only or CAVs only since CVs/CAVs are more conducive to obtaining traffic information and implementing vehicle control strategies.

\subsection{Game Theory-based Modeling}
The key to modeling with game theory is to identify three basic elements in decision-making problems, i.e., players, strategies, and payoff functions. However, identifying these elements can be challenging due to the complex nature of intersections in urban environments. The set of players can include a variety of traffic participants, such as HVs, CAVs, and pedestrians. The set of strategies can be represented by the various choices each participant can make, such as selecting a movement direction or a speed profile. The payoff function can be defined based on various factors, i.e., vehicle delay, fuel consumption, and safety. Additionally, various game theory models, such as non-cooperative and cooperative games, can be employed to analyze interactions among traffic participants at the intersection. Identifying and applying these elements and models can provide valuable insights for intersection management and can improve traffic flow efficiency and safety.

\subsubsection{Players}
Existing studies usually make two common assumptions related to the rationality and intelligence of players \cite{myerson1997game}. Rational players adopt strategies that maximize their expected payoff. If an intelligent player knows everything about the game, the player can make informed decisions regarding his or her next moves. Table \ref{table1} summarizes the literature concerning the categories and number of players involved during intersection management. The number of players in a game directly affects the choices of game theory models. Although the two-player game has been studied more extensively than other games with multiple players \cite{kamhoua2021game}, the two-player game appears to be insufficient in behavioral modeling at intersections where multiple players, such as vehicles and pedestrians, compete for the right-of-way. Therefore, in addition to the two-player matrix game, it is worthwhile to investigate multi-player games to better represent the diverse interactions at intersections. Note that individual vehicles are widely applied as players in game-theoretic modeling at intersections, and it is generally assumed that each vehicle takes the optimal action to maximize its expected payoff.  In situations where pedestrian signals are not present, pedestrians and motor vehicles also compete for right-of-way. Some researchers opt to consider players who represent a collective, such as phase, intersection, and platoon of vehicles, whose objective is to maximize the collective payoff for all the vehicles encompassed within the aggregation. 

\begin{table*}[!t]
  \caption{Players at the intersection games}
  \label{table1}
  \centering
  \scalebox{1}{
    \begin{tabular}{|c|c|c|}
      \hline
      \multirow{2}*{Players}                                              & \multicolumn{2}{c|}{Number of players}                                                                                                                                                                                                                                                                                                                                                                                                                                                                                                                                                                                                                                                                                                                                                                                                                        \\
      \cline{2-3}
      \multirow{2}*{}                                                  & Two-player                                                                                                                                                                                                                                                                                                                                                                                                                                                                                             & Multiple ($2^+$) players                                                                                                                                                                                                                                                                                                                                       \\\hline
      Vehicle/Driver                                                   & \cite{li2018game, zhang2014analysis,elhenawy2015intersection,cheng2018speed,chen2020conflict,jin2020game,tian2018adaptive,zohdy2012game,yang2016cooperative,fan2014characteristics,zhang2012dilemma,wang2011dirty,lemmer2020driver,hang2022driving,qiu2011non,bouderba2019evolutionary,cai2021game,qi2014game,liu2020game,li2022human,baz2020intersection,2018Intersection,Chen2017Modeling,2020Driver,lin2020comparative,2018Reliable,Li2021safe,liu2022three,Zhu2022,Yang2020,Wang2021,Rahmati2021,Arbis2016,lu2023game,jia2023interactive} & \cite{pruekprasert2019decision,li2020game,isukapati2017accommodating,sayin2018information,wang2021game,2020Driver,Molinari2018,Xu2019,Vasirani2012,Cheng2019,Cabri2021,Carlino2013,Schepperle2008,hang2022decision,suriyarachchi2022gameopt,chandra2022gameplan,Molinari2019,Tian2022,Levin2015,Liu2020,Wang2020,Philippe2019,Schwarting2019,Sarkar2021,yuan2021deep,Rey2021,lu2023game} \\\hline
      Pedestrian                                                       & \cite{wang2011dirty,qiu2011non,Zhu2022,Yang2020}                                                                                                                                                                                                                                                                                                                                                                                                                                                        &               -                                                                                                                                                                                                                                                                                                                                      \\\hline
      Phase/Barrier group                                              & \cite{abdelghaffar2019development,Tan2010,Bui2017,Zhao2019,tan2017multi,Khanjary2013}                                                                                                                                                                                                                                                                                                                                                                                                         & \cite{lloret2016envy,abdelghaffar2019development,NamBui2018}                                                                                                                                                                                                                                                                                        \\\hline
      Intersection/Traffic operator/Signal controller                   & \cite{qi2014game,Dong2011,Raphael2017,Xia2018Adaptive,Abdoos2021,Clempner2015}                                                                                                                                                                                                                                                                                                                                                                                                                          & \cite{Xu2019,Clempner2015,CastilloGonzalez2019,Xu2013,ZhuYT2022}                                                                                                                                                                                                                                                                                    \\\hline
      Incoming link/Path/Movement/Flow/Direction/Vehicle group/Platoon & \cite{lin2021pay,Stryszowski2021,Dai2013,Adkins2019game,MaloTamayo2009}                                                                                                                                                                                                                                                                                                                                                                                                                                 & \cite{Guo2020,2015A,guo2019optimization,Moya2015,Guo2020many}                                                                                                                                                                                                                                                                                                   \\\hline
    \end{tabular}}
\end{table*}

\subsubsection{Strategies}
In game-theoretic modeling at intersections, players choose different strategies to maximize their expected payoffs from a finite (e.g., yield or not yield) or infinite (e.g., acceleration rate) set of strategies. The strategies can also be categorized into two main types: non-cooperative and cooperative strategies. With cooperative strategies, players try to achieve mutual benefits. In non-cooperative games, on the other hand, players take actions to maximize their own payoffs without considering the payoffs of others or the system. The most common strategies are summarized below and outlined in Table \ref{table2}. 

\begin{enumerate}[label=(\roman*)]
\item \textit{Yielding}: This strategy involves a player giving up the right-of-way to others, which is usually adopted by traffic participants to avoid collision and to maintain safety at intersections.

\item \textit{Accelerating}: The strategy involves a player increasing his/her speed to approach and/or depart from an intersection, which is commonly used by traffic participants who want to avoid waiting at red lights or to pass through the intersection quickly.

\item \textit{Decelerating}: This strategy involves players reducing their speed to stay safe or to comply with traffic regulations, which is commonly adopted by drivers who want to avoid accidents or obey traffic lights.

\item \textit{Signal control}: This strategy pertains to the traffic signal controller that controls the phase and signal timing to manage traffic flow and reduce congestion at intersections.

\end{enumerate}

In summary, players at intersections adopt a variety of strategies to optimize their payoffs. The choice of strategy depends on the players' type, the intersection circumstances, and the available information for decision-making. Game-theoretic modeling can help to understand how players interact and how their choices of strategies would affect intersection operations such as traffic flow and safety.

\begin{table*}[!t]
  \centering
  \caption{Players’ strategies}
  \label{table2}
  \centering
  \begin{tabular}{|c|p{6cm}|p{5cm}|}
    \hline
    Players                                                          & \centering Strategies
                                                                  & References                                                                                                                                                                                                                                                                                                                                                                                                                                                                                                                                                                                                    \\\hline
    \multirow{8}*{Vehicle/Driver}                                 & Vehicle operation (accelerate/maintain current speed/decelerate; speed profile; go through/turn left/turn right; heading angle; steering) & \cite{pruekprasert2019decision,li2020game,li2018game,wang2021game,zhang2014analysis,elhenawy2015intersection,cheng2018speed,chen2020conflict,jin2020game,tian2018adaptive,zohdy2012game,yang2016cooperative,lemmer2020driver,hang2022driving,cai2021game,liu2020game,li2022human,baz2020intersection,2018Intersection,2020Driver,Wang2021,Cheng2019,Cabri2021,Carlino2013,hang2022decision,Tian2022,Wang2020,Philippe2019,Schwarting2019,Sarkar2021,yuan2021deep,Abdoos2021,lu2023game,jia2023interactive} \\
    \cline{2-3}
    \multirow{5}*{}                                               & Yield/Not yield                                                                                                                           & \cite{wang2011dirty,qiu2011non,Li2021safe,liu2022three,Zhu2022,Yang2020,Rahmati2021,Arbis2016}                                                                                                                                                                                                                                                                                                                                                                         \\
    \cline{2-3}
    \multirow{5}*{}                                               & Cooperation/Defection, comply/disobey                                                                                                       & \cite{fan2014characteristics,zhang2012dilemma,bouderba2019evolutionary,qi2014game}                                                                                                                                                                                                                                                                                                                                                                                \\
    \cline{2-3}
    \multirow{5}*{}                                               & Bid for the right-of-way                                                                                                                  & \cite{isukapati2017accommodating,sayin2018information,Chen2017Modeling,lin2020comparative,2018Reliable,Molinari2018,Vasirani2012,Schepperle2008,suriyarachchi2022gameopt,chandra2022gameplan,Molinari2019,Levin2015,Liu2020,Rey2021}                                                                                                                                                                                                                                                \\
    \cline{2-3}
    \multirow{5}*{}                                               & Route choice                                                                                                                              & \cite{Xu2019}                                                                                                                                                                                                                                                                                                                                                                                                                                                     \\\hline
    Pedestrian                                                    & Wait/Cross                                                                                                                                & \cite{wang2011dirty,qiu2011non,Zhu2022,Yang2020}                                                                                                                                                                                                                                                                                                                                                                                                                  \\\hline
    \multirow{2}*{Phase/Barrier group}                            & Signal control (green light time/red light time, keep current phase/change to next phase)                                                 & \cite{lloret2016envy,abdelghaffar2019development,Tan2010,Bui2017,Zhao2019,tan2017multi,Khanjary2013,NamBui2018}                                                                                                                                                                                                                                                                                                                                         \\\hline
    \multirow{5}*{Intersection/Traffic operator/Signal controller} & Signal control (green light time/red light time, keep current phase/change to next phase)                                                 & \cite{Xu2019,Dong2011,Xia2018Adaptive,Clempner2015,CastilloGonzalez2019,Xu2013,ZhuYT2022}                                                                                                                                                                                                                                                                                                                                                                         \\
    \cline{2-3}
    \multirow{4}*{}                                               & Bid for the signal time                                                                                                                   & \cite{Raphael2017}                                                                                                                                                                                                                                                                                                                                                                                                                                                \\
    \cline{2-3}
    \multirow{4}*{}                                               & Passing permitted/Passing forbidden                                                                                                       & \cite{qi2014game}                                                                                                                                                                                                                                                                                                                                                                                                                                                 \\
    \cline{2-3}
    \multirow{4}*{}                                               & Independent/Cooperative                                                                                                                   & \cite{Abdoos2021}                                                                                                                                                                                                                                                                                                                                                                                                                                                 \\\hline
    \multirow{5}{5cm}{\tabincell{c}{Incoming link/Path/Movement/Flow/Direction                                                                                                                                                                                                                                                                                                                                                                                                                                                                                                                                                                                                    \\/Vehicle group/Platoon}} & Signal control (green light time/red light time, keep current phase/change to next phase) & \cite{lin2021pay,Dai2013,Guo2020,MaloTamayo2009,guo2019optimization,Moya2015,Guo2020many}\\
    \cline{2-3}
    \multirow{4}{5cm}{}                                           & Passing permitted/Passing forbidden                                                                                                       & \cite{Adkins2019game}                                                                                                                                                                                                                                                                                                                                                                                                                                             \\
    \cline{2-3}
    \multirow{4}{5cm}{}                                           & Bid for the right-of-way                                                                                                                  & \cite{2015A}                                                                                                                                                                                                                                                                                                                                                                                                                                                      \\
    \cline{2-3}
    \multirow{4}{5cm}{}                                           & Speed profile                                                                                                                             & \cite{Stryszowski2021}                                                                                                                                                                                                                                                                                                                                                                                                                                            \\\hline
  \end{tabular}
\end{table*}

\subsubsection{Payoff functions}
The payoff function is a real-valued function defined on the set of all outcomes and strategy profiles. The payoff function of each player maps the multidimensional strategy profiles into real values to capture preferences. Importantly, the player’s payoff depends not only on his or her own strategy but also on the strategies of other players \cite{2014Game}. The factors considered in the payoff functions are also diverse among traffic scenarios. For example, traffic operators and participants may be interested in safety, efficiency, comfort, and energy efficiency at intersections.

For safety, metrics such as TTC \cite{hang2022driving,hang2022decision}, time difference to collision (TDTC) \cite{wang2021game,yang2016cooperative,Cheng2019}, post-encroachment time (PET) \cite{Zhu2022}, and other surrogate safety measures (SSMs) are usually adopted, which are listed as follows:

\begin{enumerate}[label=(\roman*)]

\item \textit{Time-to-collision (TTC)} was introduced by Hayward \cite{hayward1972near} to indicate the time it takes for two vehicles to collide when the following vehicle is moving faster than the leading vehicle in the same lane. TTC can be written as

\begin{align}\label{eq1}
  TTC(t)=\frac{x_l(t)-x_f(t)}{v_l(t)-v_f(t)},\forall v_l(t)<v_f(t)
\end{align}

\noindent where $x_l(t)$ and $v_l(t)$ are the location and velocity of the leading vehicle, respectively, and $x_f(t)$ and $v_f(t)$ pertain to the following vehicle. $t$ is the current timestamp.

\item \textit{Time different to collision (TDTC)} in \eqref{eq2} was proposed by \cite{zhang2012pedestrian} representing the time difference between the two vehicles arriving at the conflict point.

\begin{align}
  \begin{split}\label{eq2}
    TDTC(t)_{i,j}=\vert (\sqrt{(\frac{v_i(t)}{a_i(t)})^2+\frac{2L_i(t)}{a_i(t)}}-
    \frac{v_i(t)}{a_i(t)})-\\(\sqrt{(\frac{v_j(t)}{a_j(t)})^2+\frac{2L_j(t)}{a_j(t)}}
    -\frac{v_j(t)}{a_j(t)})\vert,\\ v_{min}\leq v_i(t),v_j(t)\leq v_{max}
  \end{split}
\end{align}

\noindent where $v_i(t)$ and $a_i(t)$ are the velocity and acceleration at time $t$, respectively, and $L_i(t)$ is the distance from the current position of vehicle $i$ to the conflict point. Similarly, $v_j(t)$,
$a_j(t)$, and $L_j(t)$ pertain to vehicle $j$. $v_{min}$ and $v_{max}$ are the minimum velocity and maximum velocity, respectively.

\item \textit{Post-encroachment time (PET)} is defined by  \cite{varhelyi1998drivers}, which is the time elapsed between the moment that the first vehicle leaves the virtual conflict zone and the moment
the second vehicle reaches it.

\begin{align}\label{eq3}
  PET = \vert \frac{L_i}{v_i}-\frac{L_j}{v_j}\vert
\end{align}

\noindent where $i$ and $j$ are the indexes of vehicles. $L_j$ and $L_j$ represent the distance from their position to the conflict point. $v_i$ and $v_j$ are the velocity of vehicle $i$ and vehicle $j$, respectively.

\item \textit{Time advantage (TAdv)} describes situations where two road users pass a common spatial area at different times to avoid a collision \cite{cheng2018speed,abdelghaffar2019novel}. TAdv serves as a measure to evaluate PET, with the calculation method delineated as follows,

\begin{align}\label{eq4}
  TAdv(t)=\vert \frac{L_i(t)}{v_i(t)}-\frac{L_j(t)}{v_j(t)}\vert<T_M
\end{align}

\noindent where $i$ and $j$ are the vehicle indexes, and $L_i(t)$, $L_j(t)$ denote the current distance from their position to the point of conflict. $v_i(t)$ and $v_j(t)$ are the current velocities of vehicle $i$ and vehicle $j$, respectively. $T_M$ is the time threshold of safety.

\end{enumerate}

For traffic efficiency, individual vehicles focus on microscopic information, such as velocity, delay, and waiting time; traffic operators, on the other hand, concern more about macro-level metrics such as queue length, and throughput from an aggregated/systemic perspective.

\textit{Velocity}-based metrics are usually expressed in two ways. As shown in \eqref{eq5}, the first way involves quantifying the change of speed $\Delta v_i(t)$ (i.e., acceleration or deceleration rate) of vehicle $i$, \cite{yang2016cooperative}. The second way in \eqref{eq6} is the absolute value or square of the difference between the vehicle’s current speed and the speed limit or desired speed \cite{pruekprasert2019decision}.

\begin{align}\label{eq5}
  \Delta v_i(t)=v_i(t+1)-v_i(t)
\end{align}

\begin{align}\label{eq6}
  \varphi_i(t)=\left\{\begin{matrix}
                        v_i(t+1)-v_i(t) \\
                        \vert v_i(t)-v^e\vert
                      \end{matrix}\right.
\end{align}

\noindent where $v_i(t+1)$ and $v_i(t)$ are the velocities of vehicle $i$ at time $t$ and time $t+1$, respectively, and $v^e$ is the speed limit or desired speed.

\textit{Delay} is considered as the difference between normal travel time on a roadway and the estimated travel time through a work zone. Vehicle delay may be accumulated by the increased travel distance and/or reduced speed, insufficient capacity, and temporary stoppage of the traffic flow \cite{wolshon2016traffic}. Many metrics such as the \textit{average total delay} and the \textit{average stopped delay} are widely used in payoff functions \cite{lin2021pay,lloret2016envy, Zhao2019, abdelghaffar2019novel}.

\textit{Queue length} is a common measure used to evaluate the efficiency of intersections. \textit{Average queue length} is a conventional indicator, which is the quantities of vehicles in the queue per lane per time interval \cite{hao2014cycle, Zhao2019}. \textit{Estimated queue length} is another frequent metric that has two different calculation methods. The first method calculates queue length by accumulating arrivals and departures at intersections, following input-output models. The second one analyzes the formation and dissipation of queues using the shockwave theory \cite{ban2011real}.

With respect to driving/riding comfort, the focus is typically on the smoothness of vehicle control. Efforts to enhance comfort often focus on achieving a smooth trajectory. Techniques include penalizing impractical acceleration/deceleration \cite{Molinari2018}, applying filters to omit abrupt changes in acceleration/deceleration \cite{li2022human}, and 
incorporating jerk minimization in optimization models \cite{chen2020conflict, hang2022decision}. The commonly-used metric is jerk $jerk_i(t)$ in \eqref{eq7}, which is the first-order derivative of acceleration.

\begin{align}\label{eq7}
  jerk_i(t) = \frac{\mathrm{d}a_i(t)}{\mathrm{d}t}
\end{align}

\noindent where $a_i (t)$ is the acceleration of vehicle $i$ at time $t$.

Fuel (energy) consumption and emissions are crucial for both individual drivers and traffic operators, as they are directly related to travel costs and environmental sustainability. Models such as the Comprehensive Modal Emissions Model (CMEM) \cite{barth1996modal}, Virginia Tech Microscopic (VT-Micro) model \cite{rakha2004development} and Motor Vehicle Emission Simulator (MOVES) model \cite{wang2020movestar} are generally adopted in game-theoretic modeling. These models provide fine-grained estimates of fuel consumption and emissions, enabling decision-makers to manage traffic flow while minimizing environmental impacts.

\begin{figure}[!t]
  \centering
  \includegraphics[width=0.4\textwidth]{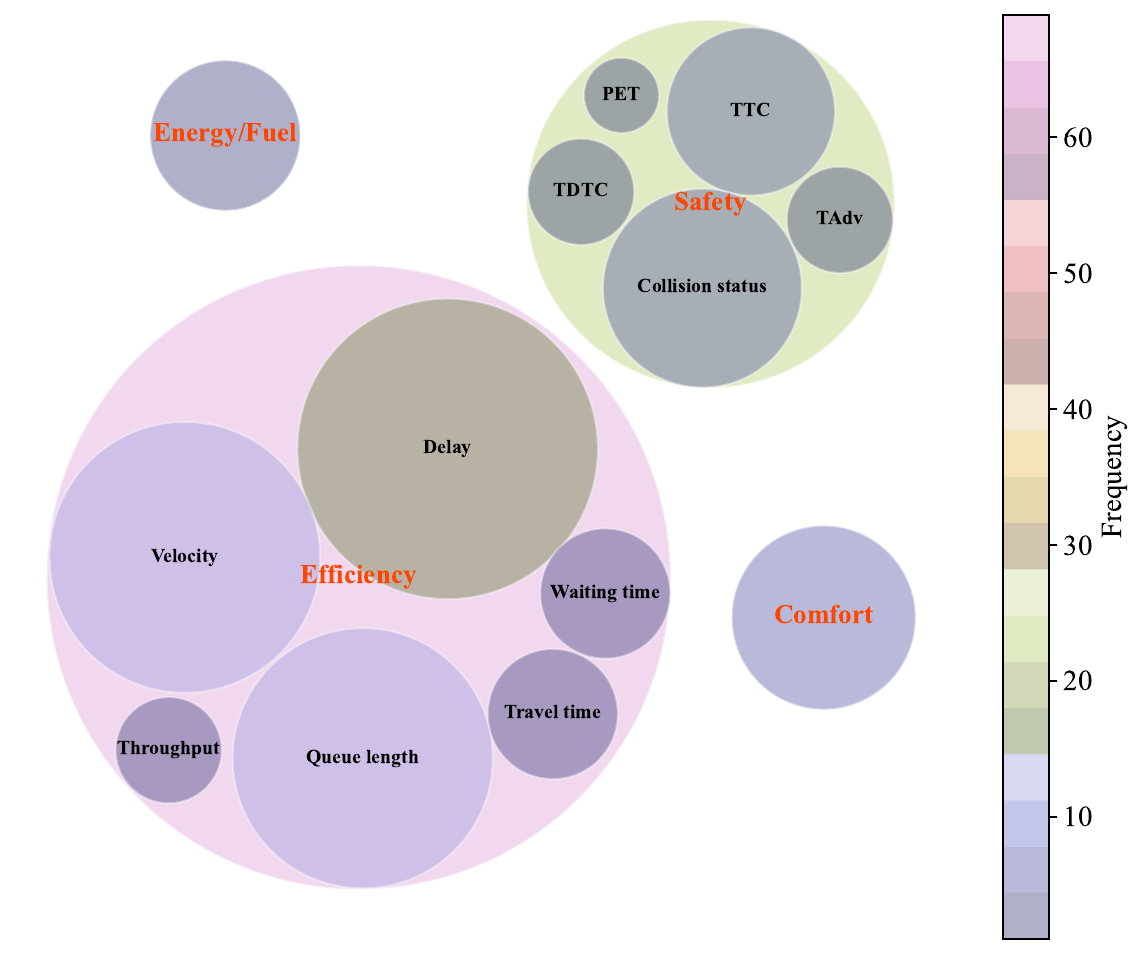}
  \caption{Factors considered in payoff functions}
  \label{fig4}
\end{figure}

Fig. \ref{fig4} visualizes the frequency of various factors used in constructing payoff functions. We employed both color and the range of circles to differentiate the frequency of various factors in determining the payoff function. Specifically, a larger range circle represents that the factor is prevalently adopted, and more detailed frequencies can be observed through different colors. According to the statistics, traffic efficiency is the most concerned, which is usually characterized by delay, queue length, and velocity. For safety, conflict-related factors such as $TTC$, and $TDTC$ are commonly used, mostly in the cases of multi-vehicle cooperation and/or un-signalized intersections. Note that only a few studies focus on comfort and energy consumption, implying that passing through intersections safely and as fast as possible is still the most urgent demand.

\subsection{Classifications of Game-theoretic Models}
Game theory is a powerful tool for modeling and analyzing complex interactions among traffic participants in traffic systems. There are various ways to classify game theory models according to different criteria, such as whether players will negotiate a binding contract and whether all information is common knowledge for players. In general, game theory models can be classified into non-cooperative games and cooperative games, which are distinguished by considering whether a coalition is formed. Non-cooperative games investigate the decision-making of individual players, while cooperative games focus on the joint actions of groups of players \cite{2014Game}. In addition, based on the amount and/or type of available information, games can also be classified into complete information games and incomplete information games. The subsequent section presents the game theory models commonly adopted in intersection management. Table \ref{table3} illustrates the applications of these game theory models and showcases their relevance and usefulness in the context of intersection management.


\begin{table*}[]
\caption{Game theory models and solutions}
\label{table3}
\centering
\begin{tabular}{|cc|c|p{4cm}|}
\hline
\multicolumn{2}{|c|}{Game type}                                                                            & Solutions                                       & References                                                                                                                                                                                                                                                      \\ \hline
\multicolumn{1}{|c|}{\multirow{6}{*}{Cooperative games}}      & \multirow{2}{*}{Transferable utility game} & Pareto efficiency                               & \cite{lin2021pay,ZhuYT2022,lloret2016envy}                                                                                                                                                                                                                    \\ \cline{3-4} 
\multicolumn{1}{|c|}{}                                        &                                            & Shapley value                                   & \cite{tan2017multi}                                                                                                                                                                                                                                 \\ \cline{2-4} 
\multicolumn{1}{|c|}{}                                        & \multirow{2}{*}{Bargaining game}           & Pareto efficiency                               & \cite{wang2021game}                                                                                                                                                                                                                            \\ \cline{3-4} 
\multicolumn{1}{|c|}{}                                        &                                            & Nash bargaining solution                        & \cite{abdelghaffar2019development,Tan2010}                                                                                                                                                                                                     \\ \cline{2-4} 
\multicolumn{1}{|c|}{}                                        & Coalition formation game                   & Merge and split rule                            & \cite{NamBui2018}                                                                                                                                                                                                                              \\ \cline{2-4} 
\multicolumn{1}{|c|}{}                                        & Fuzzy coalitional game                     & Fuzzy Shapley method                            & \cite{hang2022decision}                                                                                                                                                                                                                                \\ \hline
\multicolumn{1}{|c|}{\multirow{18}{*}{Non-cooperative games}} & \multirow{5}{*}{Auction-based game}        &                       \multirow{2}{*}{-}                          & \cite{isukapati2017accommodating,Molinari2018,Vasirani2012,Cabri2021,Schepperle2008,Molinari2019,Liu2020,suriyarachchi2022gameopt,chandra2022gameplan,Rey2021}                                                                                                \\ \cline{3-4} 
\multicolumn{1}{|c|}{}                                        &                                            & First-price auction                             & \cite{Raphael2017,Levin2015,lin2020comparative,2015A}                                                                                                                                                                                             \\ \cline{3-4} 
\multicolumn{1}{|c|}{}                                        &                                            & Second-price auction                            & \cite{lin2020comparative,Carlino2013}                                                                                                                                                                                                             \\ \cline{3-4} 
\multicolumn{1}{|c|}{}                                        &                                            & Vickrey-Clarke-Groove mechanism                 & \cite{sayin2018information}                                                                                                                                                                                                                    \\ \cline{2-4} 
\multicolumn{1}{|c|}{}                                        & \multirow{6}{*}{N-player normal form game} & Pareto improvement solution of Nash equilibrium & \cite{Dong2011}                                                                                                                                                                                                                                \\ \cline{3-4} 
\multicolumn{1}{|c|}{}                                        &                                            & \multirow{3}{*}{Nash equilibrium}                                & \cite{chen2020conflict,zohdy2012game,qiu2011non,qi2014game,2018Intersection,Chen2017Modeling,2018Reliable,Li2021safe,liu2022three,Rahmati2021,Arbis2016,CastilloGonzalez2019,Dai2013,elhenawy2015intersection,Wang2021,Xia2018Adaptive,Guo2020many,Cheng2019} \\ \cline{3-4} 
\multicolumn{1}{|c|}{}                                        &                                            & Pareto efficiency                               & \cite{Zhao2019,Abdoos2021, MaloTamayo2009}                                                                                                                                                                                                     \\ \cline{3-4} 
\multicolumn{1}{|c|}{}                                        &                                            & -                          & \cite{zhang2012dilemma,bouderba2019evolutionary}                                                                                                                                                                                               \\ \cline{2-4}
\multicolumn{1}{|c|}{}                                        & \multirow{2}{*}{Sequential game}                       & Nash equilibrium                                & \cite{li2020game,jia2023interactive,pruekprasert2019decision}      \\ \cline{3-4}       \multicolumn{1}{|c|}{}                                        &                                            & Stackelberg-Nash equilibrium                                   & \cite{Bui2017,Schwarting2019,Clempner2015,Moya2015,lu2023game}                                                                                                            \\ \cline{2-4} 
\multicolumn{1}{|c|}{}                                        & \multirow{2}{*}{Level-k model}             &                  -                               & \cite{li2018game,jin2020game,tian2018adaptive,Tian2022,Wang2020,yuan2021deep}                                                                                                                                                                     \\ \cline{3-4} 
\multicolumn{1}{|c|}{}                                        &                                            & Nash equilibrium                                & \cite{Sarkar2021}      
                                                                              \\  
\cline{2-4} 
\multicolumn{1}{|c|}{}                                        & Differential game                          & Nash equilibrium and Stackelberg equilibrium    & \cite{hang2022driving}                                  
                                                                              \\ 
\cline{2-4} 
\multicolumn{1}{|c|}{}                                        & \multirow{2}{*}{Others}                    & \multirow{2}{*}{Nash equilibrium/Pareto efficiency}    &  \cite{Khanjary2013,zhang2014analysis,yang2016cooperative,Bui2017,cheng2018speed,Xu2019,liu2020game,li2022human}                          
                                                                        \\  \hline                      
\end{tabular}
\end{table*}

Non-cooperative \textit{V.S.} Cooperative:

\subsubsection{Non-cooperative games}
Non-cooperative (strategic) game is a type of game in which individual players take strategies and in which the outcome of the game is described by the strategy taken by each player, along with the corresponding payoff for each player \cite{barkley2019economics}. In non-cooperative games, players do not form coalitions or cooperate directly with one another. Instead, they take strategy independently to maximize their individual payoffs.  Note that ``non-cooperative" does not mean there is no cooperation among players. Instead, the game focuses on the individual strategies and payoffs, and it analyzes the joint strategies without considering the possibility of some players forming coalitions and transferring their payoff within the coalition \cite{kamhoua2021game}. Non-cooperative games are widely used in modeling right-of-way competition at intersections, such as signal control, pedestrian crossing, and passing sequence decision-making of CAVs. Some commonly used non-cooperative games for intersection management include and are not limited to $N$-player normal form games, \textcolor{red}{L}evel-$k$ thinking, and Stackelberg games.

\begin{enumerate}[label=(\roman*)]
\item \textit{$N$-player normal form games}: The game describes all possible combinations of strategies and the corresponding payoffs in a matrix for two players or multiple matrices for more than two players. Players aiming to maximize their expected utilities may take a pure strategy, identified as a deterministic single strategy. Alternatively, they may resort to a mixed strategy, which assigns a probability distribution to the possible pure strategies \cite{kamhoua2021game}. A finite $N$-player normal form game can be formulated by using a tuple $\langle \mathcal{N}, \mathcal{A}, a \rangle $, where:

$\bullet \mathcal{N} =\{1,2,3, \ldots ,N\}$ is the finite set of $N$-players, indexed by $i$.

$\bullet \mathcal{A} = \mathcal{A}_1\times \mathcal{A}_2 \times \ldots \times \mathcal{A}_N $ represents the joint strategy set of players, where $\mathcal{A}_i$ is the finite strategy set available to player $i$ and $a=(a_1,a_2,\ldots,a_N) \in \mathcal{A}$ is a strategy profile with $a_i \in \mathcal{A}_i$.

$\bullet u=(u_1,u_2, \ldots, u_N)$, $u_i: \mathcal{A \to \mathbb{R} }$ is the payoff function for player $i$, which maps the strategy profile $a$ to a real value. The payoff of player $i$ is dependent not only on his or her own strategy but also on strategies taken by others. Therefore, the utility function is defined over the space of $\mathcal{A}$ instead of $\mathcal{A}_i$.

\item \textit{Level-$k$ thinking}: Level-$k$ thinking assumes that players take their strategies based on predictions about the strategies of others, thus we can categorize players by the ``depth" of their strategic thought, a.k.a., ``Levels" \cite{Tian2022}. ``Level-zero" players take their strategies without considering the strategies of others. ``Level-$k$" behave under the assumption that their fellow players are at ``Level-$k-$1". They then adopt the optimal response to the strategies of the ``Level-$k-$1" players. In \cite{Tian2022}, driving aggressiveness, encompassing both conservative vehicles and aggressive vehicles, was utilized as a criterion to demonstrate the ``Level".

\item \textit{Stackelberg games}: The Stackelberg game depicts the process in perfect information where one player is the leader who takes a strategy first. The other players are followers who observe the leader's strategy and then choose their strategies accordingly \cite{kamhoua2021game}. In the context of intersection signal control, the signal phase with emergency vehicles can be deemed as a leader who first determines its phase timing and the other phases may be considered as followers \cite{Bui2017}.

\item \textit{Auction-based game}: An auction is a mechanism that allocates a set of goods to bidders based on their announced bids. Today, auctions are pervasive in e-commerce, e-business transactions, and many other web-based applications \cite{2014Game}. With the evolution of V2X technology, they have also found applications in intersection management \cite{Schepperle2008, chandra2022gameplan, Rey2021}. Common types of auctions include the English auction, Dutch auction, first-price auction, and second-price auction. In subsequent sections on incomplete information games, we will delve into those auction types frequently used in intersection management that are incentive-compatible.
\item \textit{Differential game}: Differential games are grounded in a mathematical theory that addresses conflict problems modeled as games where players' states continuously vary over time \cite{Quincampoix2012}. In such games, the potential actions of players are modeled and analyzed using differential equations that encompass control vectors manipulated by players \cite{mkiramweni2019survey}. The inherent ability of differential games to handle continuous time challenges, combined with their resemblance to control problems featuring multiple controllers with different objectives, makes them aptly suited for intersection management, where traffic states are continuously evolving.

\end{enumerate}

\subsubsection{Cooperative games}
The cooperative (coalitional) game is a type of game in which players can negotiate binding contracts that allow them to plan joint strategies, and no players can do worse by cooperating\cite{barkley2019economics,NamBui2018}. In cooperative games, a group of rational players cooperates to reach a shared goal, in which the most crucial step is the allocation of the payoff among the players. The mathematical definition of a cooperative game can be represented using an ordered pair $\langle N,v \rangle$, where $N$ is a finite set of players and $v=2^N\to \mathbb{R}$ is the characteristic function, which describes the value created by the group of players. It conforms to the $v(\phi )=0$, signifying that the collective payoff of an empty coalition is zero (0) across all subsets of $N$ \cite{driessen2013cooperative}. Typical cooperative games used in intersection management are transferable utility games that allow a group of vehicles to ``purchase" the travel time from others \cite{lin2021pay}, and bargaining games to optimize the signal timing by modeling signal phases as players \cite{abdelghaffar2019development}.

\begin{enumerate}[label=(\roman*)]
\item \textit{Transferable utility games}: In a transferable utility game, players can make unlimited side payments to others. This allows them to redistribute the value of a coalition among the players in the coalition in any way as they deem appropriate \cite{Wilson2011}. The characteristic function in transferable utility games 
assigns a value of $v(S)$ to each coalition $S\subseteq N$, where $v(S)$ is the value of the coalition $S$ depicting the total amount of transferable utility that the members of $S$ can earn without any help from the player outside of $S$ \cite{2014Game}.

\item \textit{Bargaining games}: Bargaining games describe the process by which two or more players bargain toward an agreed-upon outcome \cite{maschler2020game}. The two-player bargaining game was first presented by \cite{nash1950bargaining}, where each player demands a portion of some good, if the sum of the proposals is no more than the total good, then both players obtain their demand. Otherwise, both players receive nothing \cite{mkiramweni2019survey}. A two-player bargaining game can be represented by using the ordered pair $(S,d)$, where $S \subseteq \mathbb{R}^2$ is the set of alternatives, which is nonempty, compact, and convex; $d=(d_1.d_2)\in S $ is the conflict (disagreement) point. There is an alternative $x=(x_1, x_2) \in S$ satisfying $x\gg d$, which means that if players agree on the alternative $x$, their payoffs are $x_1$ and $x_2$, respectively, if not, their payoffs are $d_1$ and $d_2$ \cite{maschler2020game}.
\end{enumerate}

Complete information \textit{V.S.} Incomplete information:

\subsubsection{Complete information games}
A game is regarded as a complete information game if the players share common knowledge of the game being played, including each player's strategy sets and payoff functions. This indicates that each player is aware of these elements, each player knows that all other players are aware, and this cycle of knowledge continues indefinitely \cite{2014Game,kamhoua2021game}. In traffic scenarios, particularly at intersections, complete information gaming is an overly idealized method of modeling, as traffic participants typically prefer to maintain the privacy of their trip information. Furthermore, obtaining information about all users poses a significant challenge due to communication constraints.

\subsubsection{Incomplete information games}
A game is regarded as an incomplete information (also known as asymmetric information) game when, as the players are ready to take a strategy, at least one player possesses private information about the game that the other players may not know \cite{2014Game}. In such scenarios, players must make decisions based on the available information and their beliefs about the undisclosed information. To address the information asymmetry in incomplete information games, mechanism design is commonly used to reveal the real information of the players \cite{2014Game}. Mechanism design involves designing a game that incentivizes players to reveal their private information truthfully, leading to a more efficient outcome. In mechanism design, the designer chooses a mechanism that maximizes the social welfare subject to some constraints, such as incentive compatibility, individual rationality, and budget balance. Incentive compatibility is an effective tool for revealing the real information players possess, with two main types: (1) For dominant strategy incentive compatibility (DSIC), truth revelation is the optimal response for each player, regardless of the information reported by others; (2) for Bayesian incentive compatibility (BIC). truth revelation is the optimal response for each player in expectation of the information of other players \cite{2014Game}. Incentive compatibility can be achieved by the following methods.

\begin{enumerate}[label=(\roman*)]
\item \textit{Second-price auction (Vickrey auction)}: A second-price auction is a sealed-bid auction where the highest bidder emerges as the winner but only pays a price equivalent to the second-highest bid \cite{2014Game}.

\item \textit{Vickrey-Clarke-Groves (VCG) mechanism}: The Vickrey-Clarke-Groves (VCG) mechanism is designed to eliminate incentives for misreporting by imposing on each player the cost of any distortion they cause. In the VCG mechanism, the payment for a player is set in such a way that their report cannot influence the total payoff for the rest of the players \cite{milgrom2004putting}. This approach encourages truthful reporting by aligning the individual player's interests with the overall welfare of the group.
\end{enumerate}

Research related to incomplete information environments at intersections is comparatively limited, with key studies outlined below: Initially, the second-price auction was applied to design a reservation policy for managing the passing sequence of conflicting zone movements at intersections \cite{Schepperle2008,Carlino2013}. Subsequently, the VCG mechanism was designed to reveal true private information from players, such as the urgency level, to schedule the vehicle passing sequence at intersections \cite{sayin2018information}. More recently, an online incentive-compatible mechanism was proposed to prevent players from misreporting their delay costs, notably, it can be implemented in dynamic traffic environments \cite{Rey2021}. Sponsored search auction was employed as a solution for encouraging players to report truthful information to prevent conflicts and deadlocks in the mixed traffic flow comprising HVs and CAVs \cite{chandra2022gameplan}. Following this, a sponsored search auction was extended for application in multi-vehicle dynamic traffic environments to mitigate overflow issues \cite{suriyarachchi2022gameopt}.

\subsection{Game solutions}
The solutions of a game represent the combinations of strategies that players may adopt, and in turn predict the possible outcomes of the game, which may correspond to multiple different solutions. The general solutions of games are summarized below, and the applications for intersections are listed in Table \ref{table3}.

\subsubsection{Nash equilibrium} 
Nash equilibrium is one of the solution concepts prevalent in non-cooperative games, and players have no incentives to deviate from their current strategies, even if they know others' strategies at Nash equilibrium \cite{nash1996non}.

Given a non-cooperative game $\langle \mathcal{N}, \mathcal{S} _i, u_i \rangle $, let $ \mathcal{N}={\{1,2,3,\ldots ,N\}}$ be the set of players, $\mathcal{S} _i$ be the possible strategy set of player $i$ and $u_i$ be the payoff function of player $i$. The strategy profile $s^*=(s_i^*,s_{-i}^*)$ where $s_{-i}^*$ is strategies of all players except player $i$ is the Nash equilibrium if

\begin{align}\label{eq8}
  u_i(s_i^*,s_{-i}^*)\geq u_i(s_i,s_{-i}^*), \forall s_i \in S_i, i \in \mathcal{N}
\end{align}

which can also be formulated as

\begin{align}\label{eq9}
  u_i(s_i^*,s_{-i}^*)=\max_{}  u_i(s_i,s_{-i}^*), \forall i \in \mathcal{N}
\end{align}

In short, each player's Nash equilibrium is the best response to the other players' Nash equilibrium \cite{2014Game}.

\subsubsection{Pareto efficiency}
Pareto efficiency (Pareto optimality) is an economic state where resources cannot be reallocated to make one individual better off without making at least one individual worse off \cite{2014Game}. Pareto efficiency can be expressed mathematically as

The strategy profile $s$ is Pareto efficient if there is no other strategy profile $s'$ such that the payoff $u_i(s') \geq u_i(s) $ for every player $i$ and $u_j(s') \geq u_j(s')$ for some player $j$.

\subsubsection{Shapley value}
The Shapley value is a popular single-valued solution concept for cooperative games, which provides a way to allocate total benefits to players in a coalition \cite{maschler2020game}. The Shapley value gives player $i$ in the coalition a share of

\begin{align}\label{eq10}
  \phi _i(N,v)=\sum_{S \subseteq N\setminus {i} }  \frac{\left\lvert S\right\rvert ! (n-\left\lvert S\right\rvert-1)!}{n!}(v(S\cup {i})-v(S)) 
\end{align}

\noindent in which $n$ is the total number of players, $\left\lvert S\right\rvert$ is the cardinality of coalition $S$, and $v(S)$ is the characteristic function indicating the worth of coalition $S$.

\subsubsection{Core}
In a transferable utility cooperative game, the core is the set of payoff allocations that are stable against deviations by any coalition of players, which ensures that no coalition has an incentive to break away and form a new coalition to obtain a more favorable payoff. Given a cooperative game $\langle N,v \rangle$, the core $C(N,v)$ can be represented as

\begin{align}\label{eq11}
\begin{split}
  C(N,v) = \{ \textbf{x}={x_1,x_2,x_3,\ldots,x_n}\mid \sum_{i \in N}{x_i=v(N)},\\ and\; \sum_{i \in S}{x_i\geq v(S)} \;for\; all\; S\subset N \}
\end{split}
\end{align}

\noindent where $N$ is a finite set of players, $v$ is the characteristic function, $S$ represents a coalition and $\textbf{x}=\{x_1,x_2,x_3,\ldots,x_n\}$ is a payoff vector \cite{luo2022core}.

\section{MULTI-AGENT REINFORCEMENT LEARNING AT INTERSECTIONS}
MARL has a significant connection to game theory, where players select actions to maximize payoffs in the presence of other payoff-maximizing players \cite{bowling2002multiagent}. A multi-agent system (MAS) is an extension of agent technology where a group of loosely connected autonomous agents act in an environment to achieve a common goal, which can be done either by cooperating or competing, sharing or not sharing knowledge with each other \cite{balaji2010introduction}. Recently, the requirement for adaptive MAS, combined with the imperative to examine complicated interactions among agents, has spurred the advancement of the MARL field. This is built on two fundamental pillars: the RL performed within AI and the interdisciplinary work on game theory \cite{nowe2012game}. Game theory models MAS as a multi-player game and can provide a rational strategy for each player in a game \cite{el2013multiagent}.
\subsection{A Brief Overview of Multi-agent Reinforcement Learning}
The single-agent reinforcement learning process can be formulated using a Markov decision process (MDP) \cite{yang2020overview}. The single agent will find the optimal policy through a trial-and-error process, which can maximize a possibly delayed reward in a stochastic stationary environment \cite{nowe2012game}. In the single-agent scenario, each agent solves the sequential decision-making problem by trial and error, while this comes with the difficulty of defining an optimal learning goal for multiple agents \cite{busoniu2008comprehensive}. Agents have to simultaneously interact with the environment and others, and the reward function is formulated by all agent's joint actions, making this problem even more challenging in this dynamic case. With the development in recent decades, the MARL has achieved remarkable accomplishments in many fields, such as StarCraft II \cite{vinyals2017starcraft}, Atari \cite{terry2021pettingzoo}, and Robotics \cite{zhang2021multi}, and it has also made ripples in transportation \cite{bazzan2009opportunities}.

The decision-making process in multi-agent cases is usually modeled by the stochastic game in \textbf{Definition 1} \cite{shapley1953stochastic,yang2020overview}.

\textbf{Definition 1 (Stochastic game)} A stochastic game is the extension of MDP in the multi-agent case, which can be defined by a tuple $\langle \mathcal{N}, \mathcal{S}, \{\mathcal{A}^i\}_{i\in \mathcal{N}}, \mathcal{P},  \{\mathcal{R}^i\}_{i\in \mathcal{N}},\gamma \rangle$. \\
Where\\
\begin{itemize}
  \item{$\mathcal{N}$: the number of agents, $\mathcal{N}=\{1,2, \cdots, N\}$, $\mathcal N>1$}
  \item{$\mathcal{S}$: the state space shared by all agents}
  \item{$\mathcal{A}^i$: the action space of agent $i$. Let $\mathbb{A}$:= $\mathcal {A}^1\times \mathcal {A}^2 \times \mathcal {A}^3 \times \cdots \times \mathcal {A}^N$}
  \item{$\mathcal{P} :\mathcal{S} \times \mathcal{A}^i\to \Delta (\mathcal{S})$: the transition probability from any state $s \in \mathcal{S}$ to state $s' \in \mathcal{S}$ for any joint actions $a \in \mathcal{A}$}
  \item{$\mathcal{R}^i:\mathcal{S} \times \mathcal{A} \times \mathcal{S} \to \mathbb{R}$: the reward function that returns a scalar value to the agent $i$ for a transition from $(s, a)$ to $s'$}
  \item{$\gamma \in [0,1]$ is the discount factor}
\end{itemize}

At each time step $t$, given the state $s_t$, the agent $ \in \mathcal{N} $ will execute the action $a_t^i$, simultaneously with others. The joint actions of the agents make the system transit to the next state $s_{t+1}\sim P(\cdot | s_t,a_t)$, and rewards of the agent $i$ by $R^i(s_t,a_t,s_{t+1})$. The agent $i$ aims at optimizing its long-term reward in \eqref{eq101} by finding the policy $\pi ^i:\mathcal{S} \to \Delta (\mathcal{A} ^i)$ such that $a^i \sim \pi ^i(\cdot | s_t)$. The common shorthand notation $-i=\mathcal{N}\setminus {i}$ is used to represent the set of other agents. Therefore, the value function $V^i:\mathcal{S} \to \mathbb{R} $ of agent $i$ is transformed into the joint policy $\pi: \mathcal{S} \to \Delta (\mathcal{A})$ defined as $\pi (a |s):=\prod _{i \in \mathcal{N}}\pi^i(a^i|s)$. Given the specific joint policy $\pi$ and state $s \in \mathcal{S}$.\\
\begin{align}\label{eq101}
  V_{\pi^i,\pi^{-i}}^i(s):=\mathbb{E}[\sum_{t\geq 0}\gamma ^tR^i(s_t,a_t,s_{t+1})| a_t^i\sim \pi^i(\cdot|s_t),s_0=s]
\end{align}
The optimal policy is\\
\begin{align}\label{eq102}
  \pi _i ^*(a_i|s,\pi_{-i})=\underset{\pi _i}{{\arg\max}} \,V_i^{\pi _i,\pi _{-i}}(s)
\end{align}
We recommend that interested readers refer to \cite{yang2020overview} and \cite{zhang2021multi} for a detailed introduction to the MARL.

\subsection{Benchmarks and Simulation Platforms}
The opportunities of multi-agent learning in intersection management pointed out by Dresner and Stone have attracted and encouraged an increasing number of scholars to address the challenges of MARL in intersections \cite{dresner2006multiagent}. Especially, MARL is widely considered to be promising for traffic signal control (TSC) in large-scale road networks. The topic of MARL applications for intersection management has attracted increasing attention from researchers.

In the domain of intersection management, the application of MARL often entails performance comparisons with conventionally recognized methods to underscore their advantages. Notably, several high-performing AI-based techniques have also gained prominence. Hence, we provide a list of potential benchmarks here to facilitate related research.
\subsubsection*{\bf Conventional methods}
\begin{itemize}
  \item{Fixed time \cite{koonce2008traffic}}
  \item{MaxPressure \cite{varaiya2013max}}
  \item{GreenWave \cite{roess2011traffic}}
  \item{Longest queue first \cite{wu2019dcl}}
  \item{SCOOT \cite{robertson1991optimizing}}
  \item{SOTL \cite{cools2013self}}
\end{itemize}
\subsubsection*{\bf AI-based methods}
\begin{itemize}
  \item{PressLight \cite{wei2019presslight}}
  \item{CoLight \cite{wei2019colight}}
  \item{MA2C \cite{chu2019multi}}
  \item{MetaLight \cite{zang2020metalight}}
  \item{IG-RL \cite{devailly2021ig}}
  \item{MARLIN-ATSC \cite{el2013multiagent}}
  \item{MPLight \cite{chen2020toward}}
  \item{AttendLight \cite{oroojlooy2020attendlight}}
\end{itemize}

MARL requires considerable training before it can be applied in practice, so a simulation platform that provides a more realistic picture of traffic operations is critical. With the rapid development of MARL technology and its widespread application at intersections, some platforms are becoming compatible with multi-agent training. This section lists some prevailing platforms.\\
\begin{itemize}
  \item{SUMO \cite{lopez2018microscopic}}
  \item{Paramics \cite{cameron1996paramics}}
  \item{Aimsun \cite{casas2010traffic}}
  \item{VISSIM \cite{fellendorf2010microscopic}}
  \item{CityFlow \cite{zhang2019cityflow}}
  \item{Flow \cite{wu2021flow}}
  \item{Traffic3D \cite{garg2019traffic3d}}
  \item{SMARTS \cite{zhou2020smarts}}
  \item{highway-env \cite{highway-env}}
  \item{Movsim \cite{Movesim2010}}
  \item{Carla \cite{dosovitskiy2017carla}}
\end{itemize}

\subsection{Pros and Cons of Game Theory V.S. MARL}
Both game theory and MARL can tackle decision-making in multi-agent environments. Game theory offers an analytical method, shedding light on players' strategies and producing explainable outcomes like the Nash equilibrium and the Pareto optimality. However, its efficacy diminishes with continuous actions or an extensive number of agents while MARL is adept at generating continuous actions for numerous agents. Despite its strengths, MARL's challenges include time-consuming training, transferability issues, and lack of explainability. Recognizing these possible shortcomings, hybrid algorithms like Minimax-Q and Nash-Q have emerged, integrating game solutions with MARL to derive learning-based equilibrium \cite{bowling2000analysis}. These hybrids aim to balance the rigor of game-theoretic solutions with the dynamic learning capabilities of MARL, addressing their individual limitations while leveraging their collective strengths.

\section{Limitations and future research directions}
To facilitate the adoption of game theory in intersection management, it is essential to address the technical challenges associated with these applications and to identify potential future research directions. This section discusses some of the main challenges in using game theory for intersection management, as well as promising approaches and future research directions in this field. 

\subsection{Limitations of existing related studies}
In the previous sections, the applications of game theory for intersections were reviewed and summarized. However, despite numerous studies that have investigated various intersection environments, there are still many limitations, as listed below.

\begin{enumerate}[label=(\roman*)]
\item One significant challenge is the development of game-theoretic models in dynamic traffic environments. Traffic states at intersections can exhibit marked changes dynamically, such as the arrival of vehicles and vehicle states. Therefore, the game theory model should be able to adapt to such changes and to make real-time decisions. Additionally, the computational complexity of game-theoretic models can make them challenging to implement in real-time traffic management systems \cite{li2020game}. Thus, there is a need to develop more efficient algorithms and techniques to calculate with these models efficiently.

\item {Communication plays a crucial role in micro-level interactions among traffic participants, but some studies do not specify the degree of connectivity and intelligence required for effective communication. In general, vehicular communications rely on robust link and channel access protocols, such as IEEE 802.11p wireless access in vehicular environments (WAVE) \cite{martinez2010emergency}. It is worth noting several inherent communication issues: (a) frequent topology partitioning resulting from the high-speed mobility of traffic participants, (b) instability of long-range communication due to increased delay spread and diminished channel capacity, (c) complications arising from the use of conventional routing protocols, and (d) broadcast storm issue in high dense scenarios. Real-time performance is crucial for intersection management. Without a reliable communication network or accurate adherence to control strategies by vehicles, intersections can become highly disorganized and inefficient. Consequently, potential communication challenges need further discussion and investigation.} 

\item NmT, who are highly involved in traffic and interact frequently still receive insufficient attention in related studies, and more research is needed to address their specific needs and challenges.

\item The assumption of a complete information environment may be too idealistic in real-world scenarios, and traffic participants may misreport their private information to gain more benefits, resulting in unfair resource allocation. Thus, more research is needed to develop mechanisms to detect and to mitigate the effects of information asymmetry.

\item Existing studies have skillfully applied game theory to upper-level decision-making for autonomous driving. However, the upper-level decisions made by game theory models are not consistently well-integrated with lower-level vehicle motion planning and control.

\item Game theory is formulated under the assumption of rational players, but players' strategies can also be influenced by different characteristics. In existing studies, many utilitarian factors, such as total delay or throughput, are considered, but the real needs and preferences of traffic participants are often not adequately addressed. Heterogeneous users, with their diverse preferences and objectives, need further investigation in addition to the mere differentiation in the categories of players.
\end{enumerate}

\subsection{Future research directions}
Based on the limitations of the available literature, several gaps and opportunities for future research in game theory-based intersection management can be identified. These include:

\begin{enumerate}[label=(\roman*)]
\item For micro-level interactions between vehicles, it is crucial to address the communication requirements of traffic participants in real-time scenarios. Simulating a connected environment by adding communication systems to traffic simulation is one way to improve the realism of the simulations. Additionally, field experiments under connected conditions can provide valuable insights.

\item  The interaction among MT and NmT should receive more attention in future research. These groups are highly involved in traffic and interact frequently, but their interactions are often overlooked in existing studies.

\item Incomplete information environments are common in real-world scenarios, and mechanism design under incomplete information in a dynamic traffic environment should be highlighted. Future research can explore the interaction of traffic participants by considering incentive compatibility and other constraints.

\item Combining optimization algorithms and game theory in decision-making is a promising research direction. For instance, game theory models can be used to decide whether to adopt the planned vehicle trajectory or speed profiles obtained from system-level optimization, thereby improving the efficiency of intersections.

\item  Existing models assume rational players, but drivers' and passengers' strategies can also be influenced by their emotions and experience. Future research should consider the individual characteristics and bounded rationality of traffic participants. In addition, including more personal travel information in constructing the payoff function can make intersection management acceptable to all individuals.

\item Future research directions in this field also include the exploration of new game-theoretic models and techniques for managing CAVs at intersections, the development of multi-agent reinforcement learning approaches for adaptive intersection management, and the investigation of game-theoretic approaches for managing interactions between MT and NmT.
\end{enumerate}

Overall, game theory has the potential to revolutionize intersection management by providing more efficient and adaptive solutions to traffic control. However, significant technical challenges must be addressed, and innovative approaches must be developed to fully realize the benefits of game theory in this field.

\section{Conclusion}
In conjunction with the advancement of intelligent transportation systems, incorporating knowledge of game theory and mechanism design becomes imperative for ensuring the ground truth collected individual travel information (such as the value of time), thereby fostering fair decision-making processes. Moreover, rather than the conventional collective approach to intersection management considering all vehicles in an approach as a unified group, there is a growing emphasis on personalized guidance. Game theory becomes instrumental in this shift, offering insights into individual behaviors and their interactions.
In this paper, we provided a summary of the applications of game theory for urban intersections. First, we proposed and applied a workflow for systematic literature retrieval and analysis, yielding high-quality relevant literature. Then, numerous game-theoretic modeling approaches and corresponding possible solutions applied to intersections were reviewed and summarized. Concurrently, the widely discussed multi-agent reinforcement learning (MARL), which is inextricably linked to game theory, is briefly summarized. Finally, limitations and possible future directions in this area were pointed out. In conclusion, numerous innovative solutions for intersection management have been proposed by leveraging the advantages of game theory. Game theory excels in resolving traffic conflicts and harmonizing the interests of multiple traffic participants, making it an invaluable tool in the field of transportation management. As the intelligent transportation system continues to evolve, the strategic application of game theory for analyzing individual traffic participants' interactions can significantly contribute to establishing a harmonious transportation ecosystem.

\section*{Acknowledgments}
The work is supported by the National Natural Science Foundation of China via grant 52072316 and 52302418, the Postdoctoral International Exchange Program via grant YJ20220311, the Fundamental Research Funds for the Central Universities via grant 2682023CX047, and the Tencent-SWJTU Joint Laboratory of Intelligent Transportation Program via grant R113623H01015. Any opinions, findings, conclusions, or recommendations expressed in this paper are those of the authors and do not necessarily reflect the views of the sponsors.

\bibliographystyle{IEEEtran}
\bibliography{IEEEabrv,GT_INTERSECTION_REF}

\begin{thebibliography}{100}
\providecommand{\url}[1]{#1}
\csname url@samestyle\endcsname
\providecommand{\newblock}{\relax}
\providecommand{\bibinfo}[2]{#2}
\providecommand{\BIBentrySTDinterwordspacing}{\spaceskip=0pt\relax}
\providecommand{\BIBentryALTinterwordstretchfactor}{4}
\providecommand{\BIBentryALTinterwordspacing}{\spaceskip=\fontdimen2\font plus
\BIBentryALTinterwordstretchfactor\fontdimen3\font minus \fontdimen4\font\relax}
\providecommand{\BIBforeignlanguage}[2]{{%
\expandafter\ifx\csname l@#1\endcsname\relax
\typeout{** WARNING: IEEEtran.bst: No hyphenation pattern has been}%
\typeout{** loaded for the language `#1'. Using the pattern for}%
\typeout{** the default language instead.}%
\else
\language=\csname l@#1\endcsname
\fi
#2}}
\providecommand{\BIBdecl}{\relax}
\BIBdecl

\bibitem{mirheli2019consensus}
A.~Mirheli, M.~Tajalli, L.~Hajibabai, and A.~Hajbabaie, ``A consensus-based distributed trajectory control in a signal-free intersection,'' \emph{Transp. Res. Part C, Emerg. Technol.}, vol. 100, pp. 161--176, 2019, doi: {\color{blue} \href{https://doi.org/10.1016/j.trc.2019.01.004} {10.1016/j.trc.2019.01.004}}.

\bibitem{yu2018integrated}
C.~Yu, Y.~Feng, H.~X. Liu, W.~Ma, and X.~Yang, ``Integrated optimization of traffic signals and vehicle trajectories at isolated urban intersections,'' \emph{Transp. Res. Part B, Methodol.}, vol. 112, pp. 89--112, 2018, doi: {\color{blue} \href{https://doi.org/10.1016/j.trb.2018.04.007}{10.1016/j.trb.2018.04.007}}.

\bibitem{qian2019toward}
B.~Qian, H.~Zhou, F.~Lyu, J.~Li, T.~Ma, and F.~Hou, ``Toward collision-free and efficient coordination for automated vehicles at unsignalized intersection,'' \emph{{IEEE} Internet Things J.}, vol.~6, no.~6, pp. 10\,408--10\,420, 2019, doi: {\color{blue} \href{https://doi.org/10.1109/JIOT.2019.2939180}{10.1109/JIOT.2019.2939180}}.

\bibitem{dey2016vehicle}
K.~C. Dey, A.~Rayamajhi, M.~Chowdhury, P.~Bhavsar, and J.~Martin, ``Vehicle-to-vehicle (v2v) and vehicle-to-infrastructure (v2i) communication in a heterogeneous wireless network--performance evaluation,'' \emph{Transp. Res. Part C, Emerg. Technol.}, vol.~68, pp. 168--184, 2016, doi: {\color{blue} \href{https://doi.org/10.1016/j.trc.2016.03.008}{10.1016/j.trc.2016.03.008}}.

\bibitem{sun2020cooperative}
Z.~Sun, T.~Huang, and P.~Zhang, ``Cooperative decision-making for mixed traffic: A ramp merging example,'' \emph{Transp. Res. Part C, Emerg. Technol.}, vol. 120, p. 102764, 2020, doi: {\color{blue} \href{https://doi.org/10.1016/j.trc.2020.102764}{10.1016/j.trc.2020.102764}}.

\bibitem{chen2022milestones}
L.~Chen, Y.~Li, C.~Huang, B.~Li, Y.~Xing, D.~Tian, L.~Li, Z.~Hu, X.~Na, Z.~Li, S.~Teng, C.~Lv, J.~Wang, D.~Cao, N.~Zheng, and F.-Y. Wang, ``Milestones in autonomous driving and intelligent vehicles: Survey of surveys,'' \emph{IEEE Trans. Intell. Veh.}, vol.~8, no.~2, pp. 1046--1056, 2023, doi: {\color{blue} \href{ https://doi.org/10.1109/TIV.2022.3223131} {10.1109/TIV.2022.3223131}}.

\bibitem{dresner2008multiagent}
K.~Dresner and P.~Stone, ``A multiagent approach to autonomous intersection management,'' \emph{J. Artif. Intell. Res.}, vol.~31, pp. 591--656, 2008, doi: {\color{blue} \href{https://doi.org/10.1613/jair.2502}{10.1613/jair.2502}}.

\bibitem{wang2022social}
W.~Wang, L.~Wang, C.~Zhang, C.~Liu, L.~Sun \emph{et~al.}, ``Social interactions for autonomous driving: A review and perspectives,'' \emph{Found. Trends Robot.}, vol.~10, no. 3-4, pp. 198--376, 2022, doi: {\color{blue} \href{http://dx.doi.org/10.1561/2300000078} {10.1561/2300000078}}.

\bibitem{pruekprasert2019decision}
S.~Pruekprasert, X.~Zhang, J.~Dubut, C.~Huang, and M.~Kishida, ``Decision making for autonomous vehicles at unsignalized intersection in presence of malicious vehicles,'' in \emph{Proc. IEEE Intell. Transp. Syst. Conf. (ITSC)}.\hskip 1em plus 0.5em minus 0.4em\relax IEEE, 2019, pp. 2299--2304, doi: {\color{blue} \href{http://dx.doi.org/10.1109/ITSC.2019.8917132} {10.1109/ITSC.2019.8917132}}.

\bibitem{liu2020reservation}
J.~Liu, P.~Lin, and B.~Ran, ``A reservation-based coordinated transit signal priority method for bus rapid transit system with connected vehicle technologies,'' \emph{{IEEE} Intell. Transp. Syst. Mag.}, vol.~13, no.~4, pp. 17--30, 2020, doi: {\color{blue} \href{http://doi.org/10.1109/MITS.2020.3014151} {10.1109/MITS.2020.3014151}}.

\bibitem{liu2019traffic}
H.~Liu, X.-Y. Lu, and S.~E. Shladover, ``Traffic signal control by leveraging cooperative adaptive cruise control (cacc) vehicle platooning capabilities,'' \emph{Transp. Res. Part C, Emerg. Technol.}, vol. 104, pp. 390--407, 2019, doi: {\color{blue} \href{https://doi.org/10.1016/j.trc.2019.05.027} {10.1016/j.trc.2019.05.027}}.

\bibitem{guo2019joint}
Y.~Guo, J.~Ma, C.~Xiong, X.~Li, F.~Zhou, and W.~Hao, ``Joint optimization of vehicle trajectories and intersection controllers with connected automated vehicles: Combined dynamic programming and shooting heuristic approach,'' \emph{Transp. Res. Part C, Emerg. Technol.}, vol.~98, pp. 54--72, 2019, doi: {\color{blue} \href{https://doi.org/10.1016/j.trc.2018.11.010} {10.1016/j.trc.2018.11.010}}.

\bibitem{feng2018spatiotemporal}
Y.~Feng, C.~Yu, and H.~X. Liu, ``Spatiotemporal intersection control in a connected and automated vehicle environment,'' \emph{Transp. Res. Part C, Emerg. Technol.}, vol.~89, pp. 364--383, 2018, doi: {\color{blue} \href{https://doi.org/10.1016/j.trc.2018.02.001} {10.1016/j.trc.2018.02.001}}.

\bibitem{hao2018eco}
P.~Hao, G.~Wu, K.~Boriboonsomsin, and M.~J. Barth, ``Eco-approach and departure (ead) application for actuated signals in real-world traffic,'' \emph{{IEEE} Trans. Intell. Transp. Syst.}, vol.~20, no.~1, pp. 30--40, 2018, doi: {\color{blue} \href{http://doi.org/10.1109/TITS.2018.2794509} {10.1109/TITS.2018.2794509}}.

\bibitem{sun2022eco}
P.~Sun, D.~Nam, R.~Jayakrishnan, and W.~Jin, ``An eco-driving algorithm based on vehicle to infrastructure (v2i) communications for signalized intersections,'' \emph{Transp. Res. Part C, Emerg. Technol.}, vol. 144, p. 103876, 2022, doi: {\color{blue} \href{https://doi.org/10.1016/j.trc.2022.103876} {10.1016/j.trc.2022.103876}}.

\bibitem{lin2017autonomous}
P.~Lin, J.~Liu, P.~J. Jin, and B.~Ran, ``Autonomous vehicle-intersection coordination method in a connected vehicle environment,'' \emph{{IEEE} Intell. Transp. Syst. Mag.}, vol.~9, no.~4, pp. 37--47, 2017, doi: {\color{blue} \href{http://doi.org/10.1109/MITS.2017.2743167} {10.1109/MITS.2017.2743167}}.

\bibitem{yang2020eco}
H.~Yang, F.~Almutairi, and H.~Rakha, ``Eco-driving at signalized intersections: A multiple signal optimization approach,'' \emph{{IEEE} Trans. Intell. Transp. Syst.}, vol.~22, no.~5, pp. 2943--2955, 2020, doi: {\color{blue} \href{http://doi.org/10.1109/TITS.2020.2978184} {10.1109/TITS.2020.2978184}}.

\bibitem{xiong2021speed}
B.-K. Xiong and R.~Jiang, ``Speed advice for connected vehicles at an isolated signalized intersection in a mixed traffic flow considering stochasticity of human driven vehicles,'' \emph{{IEEE} Trans. Intell. Transp. Syst.}, vol.~23, no.~8, pp. 11\,261--11\,272, 2021, doi: {\color{blue} \href{https://doi.org/10.1109/TITS.2021.3102430} {10.1109/TITS.2021.3102430}}.

\bibitem{chen2021mixed}
C.~Chen, J.~Wang, Q.~Xu, J.~Wang, and K.~Li, ``Mixed platoon control of automated and human-driven vehicles at a signalized intersection: dynamical analysis and optimal control,'' \emph{Transp. Res. Part C, Emerg. Technol.}, vol. 127, p. 103138, 2021, doi: {\color{blue} \href{https://doi.org/10.1016/j.trc.2021.103138} {10.1016/j.trc.2021.103138}}.

\bibitem{li2020game}
N.~Li, Y.~Yao, I.~Kolmanovsky, E.~Atkins, and A.~R. Girard, ``Game-theoretic modeling of multi-vehicle interactions at uncontrolled intersections,'' \emph{{IEEE} Trans. Intell. Transp. Syst.}, 2020, doi: {\color{blue} \href{https://doi.org/10.1109/TITS.2020.3026160} {10.1109/TITS.2020.3026160}}.

\bibitem{hu2021constraint}
Z.~Hu, J.~Huang, D.~Yang, and Z.~Zhong, ``Constraint-tree-driven modeling and distributed robust control for multi-vehicle cooperation at unsignalized intersections,'' \emph{Transp. Res. Part C, Emerg. Technol.}, vol. 131, p. 103353, 2021, doi: {\color{blue} \href{https://doi.org/10.1016/j.trc.2021.103353} {10.1016/j.trc.2021.103353}}.

\bibitem{chen2022conflict}
C.~Chen, Q.~Xu, M.~Cai, J.~Wang, J.~Wang, and K.~Li, ``Conflict-free cooperation method for connected and automated vehicles at unsignalized intersections: Graph-based modeling and optimality analysis,'' \emph{{IEEE} Trans. Intell. Transp. Syst.}, vol.~23, no.~11, pp. 21\,897--21\,914, 2022, doi: {\color{blue} \href{https://doi.org/10.1109/TITS.2022.3182403} {10.1109/TITS.2022.3182403}}.

\bibitem{luo2023real}
J.~Luo, T.~Zhang, R.~Hao, D.~Li, C.~Chen, Z.~Na, and Q.~Zhang, ``Real-time cooperative vehicle coordination at unsignalized road intersections,'' \emph{{IEEE} Trans. Intell. Transp. Syst.}, 2023, doi: {\color{blue} \href{https://doi.org/10.1109/TITS.2023.3243940} {10.1109/TITS.2023.3243940}}.

\bibitem{huang2009cyclists}
L.~Huang and J.~Wu, ``Cyclists' path planning behavioral model at unsignalized mixed traffic intersections in china,'' \emph{{IEEE} Intell. Transp. Syst. Mag.}, vol.~1, no.~2, pp. 13--19, 2009, doi: {\color{blue} \href{https://doi.org/10.1109/MITS.2009.933859} {10.1109/MITS.2009.933859}}.

\bibitem{isukapati2017accommodating}
I.~K. Isukapati and S.~F. Smith, ``Accommodating high value-of-time drivers in market-driven traffic signal control,'' in \emph{Proc. 2017 IEEE Intell. Veh. Symp. (IV)}.\hskip 1em plus 0.5em minus 0.4em\relax Los Angeles, CA, USA: IEEE, Jul. 2017, pp. 1280--1286, doi: {\color{blue} \href{https://doi.org/10.1109/IVS.2017.7995888} {10.1109/IVS.2017.7995888}}.

\bibitem{niroumand2020joint}
R.~Niroumand, M.~Tajalli, L.~Hajibabai, and A.~Hajbabaie, ``Joint optimization of vehicle-group trajectory and signal timing: Introducing the white phase for mixed-autonomy traffic stream,'' \emph{Transp. Res. Part C, Emerg. Technol.}, vol. 116, p. 102659, 2020, doi: {\color{blue} \href{https://doi.org/10.1016/j.trc.2020.102659} {10.1016/j.trc.2020.102659}}.

\bibitem{lin2021pay}
D.~Lin and S.~E. Jabari, ``Pay for intersection priority: A free market mechanism for connected vehicles,'' \emph{{IEEE} Trans. Intell. Transp. Syst.}, 2021, doi: {\color{blue} \href{https://doi.org/10.1109/TITS.2020.3048475} {10.1109/TITS.2020.3048475}}.

\bibitem{zhao2011computational}
D.~Zhao, Y.~Dai, and Z.~Zhang, ``Computational intelligence in urban traffic signal control: A survey,'' \emph{IEEE Trans. Syst., Man, Cybern. C, Appl. Rev.}, vol.~42, no.~4, pp. 485--494, 2011, doi: {\color{blue}\href{http://dx. doi.org/10.1109/TSMCC.2011.2161577} {10.1109/TSMCC.2011.2161577}}.

\bibitem{araghi2015review}
S.~Araghi, A.~Khosravi, and D.~Creighton, ``A review on computational intelligence methods for controlling traffic signal timing,'' \emph{Expert Syst. Appl.}, vol.~42, no.~3, pp. 1538--1550, 2015, doi: {\color{blue}\href{https://doi.org/10.1016/j.eswa.2014.09.003} {10.1016/j.eswa.2014.09.003}}.

\bibitem{wang2018review}
Y.~Wang, X.~Yang, H.~Liang, and Y.~Liu, ``A review of the self-adaptive traffic signal control system based on future traffic environment,'' \emph{J. Adv. Transp.}, vol. 2018, 2018, doi: {\color{blue}\href{ https://doi.org/10.1155/2018/10961237} {10.1155/2018/10961237}}.

\bibitem{yau2017survey}
K.-L.~A. Yau, J.~Qadir, H.~L. Khoo, M.~H. Ling, and P.~Komisarczuk, ``A survey on reinforcement learning models and algorithms for traffic signal control,'' \emph{ACM Comput. Surv.}, vol.~50, no.~3, pp. 1--38, 2017, doi: {\color{blue}\href{https://doi.org/10.1145/3068287} {10.1145/3068287}}.

\bibitem{florin2015survey}
R.~Florin and S.~Olariu, ``A survey of vehicular communications for traffic signal optimization,'' \emph{Veh. Commun.}, vol.~2, no.~2, pp. 70--79, 2015, doi: {\color{blue}\href{https://doi.org/10.1016/j.vehcom.2015.03.002} {10.1016/j.vehcom.2015.03.002}}.

\bibitem{chen2015cooperative}
L.~Chen and C.~Englund, ``Cooperative intersection management: A survey,'' \emph{{IEEE} Trans. Intell. Transp. Syst.}, vol.~17, no.~2, pp. 570--586, 2015, doi: {\color{blue}\href{ https://doi.org/10.1109/TITS.2015.2471812} {10.1109/TITS.2015.2471812}}.

\bibitem{rios2016survey}
J.~Rios-Torres and A.~A. Malikopoulos, ``A survey on the coordination of connected and automated vehicles at intersections and merging at highway on-ramps,'' \emph{{IEEE} Trans. Intell. Transp. Syst.}, vol.~18, no.~5, pp. 1066--1077, 2016, doi: {\color{blue}\href{ https://doi.org/10.1109/TITS.2016.2600504} {10.1109/TITS.2016.2600504}}.

\bibitem{guo2019urban}
Q.~Guo, L.~Li, and X.~J. Ban, ``Urban traffic signal control with connected and automated vehicles: A survey,'' \emph{Transp. Res. Part C, Emerg. Technol.}, vol. 101, pp. 313--334, 2019, doi: {\color{blue}\href{ https://doi.org/10.1016/j.trc.2019.01.026} {10.1016/j.trc.2019.01.026}}.

\bibitem{khayatian2020survey}
M.~Khayatian, M.~Mehrabian, E.~Andert, R.~Dedinsky, S.~Choudhary, Y.~Lou, and A.~Shirvastava, ``A survey on intersection management of connected autonomous vehicles,'' \emph{ACM Trans. Cyber-Phys. Syst}, vol.~4, no.~4, pp. 1--27, 2020, doi: {\color{blue}\href{ https://doi.org/10.1145/3407903} {10.1145/3407903}}.

\bibitem{namazi2019intelligent}
E.~Namazi, J.~Li, and C.~Lu, ``Intelligent intersection management systems considering autonomous vehicles: A systematic literature review,'' \emph{IEEE Access}, vol.~7, pp. 91\,946--91\,965, 2019, doi: {\color{blue}\href{ https://doi.org/10.1109/ACCESS.2019.2927412} {10.1109/ACCESS.2019.2927412}}.

\bibitem{al2022signalized}
M.~Al-Turki, N.~T. Ratrout, S.~M. Rahman, and K.~J. Assi, ``Signalized intersection control in mixed autonomous and regular vehicles traffic environment--a critical review focusing on future control,'' \emph{IEEE Access}, 2022, doi: {\color{blue}\href{ https://doi.org/10.1109/ACCESS.2022.3148706} {10.1109/ACCESS.2022.3148706}}.

\bibitem{shirazi2016looking}
M.~S. Shirazi and B.~T. Morris, ``Looking at intersections: a survey of intersection monitoring, behavior and safety analysis of recent studies,'' \emph{{IEEE} Trans. Intell. Transp. Syst.}, vol.~18, no.~1, pp. 4--24, 2016, doi: {\color{blue}\href{ https://doi.org/10.1109/TITS.2016.2568920} {10.1109/TITS.2016.2568920}}.

\bibitem{zhong2020autonomous}
Z.~Zhong, M.~Nejad, and E.~E. Lee, ``Autonomous and semiautonomous intersection management: A survey,'' \emph{{IEEE} Intell. Transp. Syst. Mag.}, vol.~13, no.~2, pp. 53--70, 2020, doi: {\color{blue}\href{ https://doi.org/10.1109/MITS.2020.3014074} {10.1109/MITS.2020.3014074}}.

\bibitem{iliopoulou2022survey}
C.~Iliopoulou, K.~Kepaptsoglou, and E.~I. Vlahogianni, ``A survey on market-inspired intersection control methods for connected vehicles,'' \emph{{IEEE} Intell. Transp. Syst. Mag.}, vol.~15, no.~2, pp. 162--176, 2022, doi: {\color{blue} \href{http://dx.doi.org/10.1109/MITS.2022.3203573} {10.1109/MITS.2022.3203573}}.

\bibitem{myerson1997game}
R.~B. Myerson, \emph{Game theory: Analysis of conflict}.\hskip 1em plus 0.5em minus 0.4em\relax Harvard University Press, 1997.

\bibitem{ji2020review}
A.~Ji and D.~Levinson, ``A review of game theory models of lane changing,'' \emph{Transportmetrica A, Transport Sci.}, vol.~16, no.~3, pp. 1628--1647, 2020, doi: {\color{blue} \href{https://doi.org/10.1080/23249935.2020.1770368} {10.1080/23249935.2020.1770368}}.

\bibitem{kamhoua2021game}
C.~A. Kamhoua, C.~D. Kiekintveld, F.~Fang, and Q.~Zhu, \emph{Game theory and machine learning for cyber security}.\hskip 1em plus 0.5em minus 0.4em\relax John Wiley \& Sons, 2021.

\bibitem{ahmad2023game}
F.~Ahmad, O.~Almarri, Z.~Shah, and L.~Al-Fagih, ``Game theory applications in traffic management: A review of authority-based travel modelling,'' \emph{Travel Behav. Soc.}, vol.~32, p. 100585, 2023, doi: {\color{blue} \href{https://doi.org/10.1016/j.tbs.2023.100585} {10.1016/j.tbs.2023.100585}}.

\bibitem{zhang2019game}
Q.~Zhang, R.~Langari, H.~E. Tseng, D.~Filev, S.~Szwabowski, and S.~Coskun, ``A game theoretic model predictive controller with aggressiveness estimation for mandatory lane change,'' \emph{IEEE Trans. Intell. Veh.}, vol.~5, no.~1, pp. 75--89, 2020, doi: {\color{blue} \href{ https://doi.org/10.1109/TIV.2019.2955367} {10.1109/TIV.2019.2955367}}.

\bibitem{ji2020estimating}
A.~Ji and D.~Levinson, ``Estimating the social gap with a game theory model of lane changing,'' \emph{{IEEE} Trans. Intell. Transp. Syst.}, vol.~22, no.~10, pp. 6320--6329, 2020, doi: {\color{blue} \href{https://doi.org/10.1109/TITS.2020.2991242} {10.1109/TITS.2020.2991242}}.

\bibitem{ji2023pricing}
A.~Ji, M.~Ramezani, and D.~Levinson, ``Pricing lane changes,'' \emph{Transp. Res. Part C, Emerg. Technol.}, vol. 149, p. 104062, 2023, doi: {\color{blue} \href{https://doi.org/10.1016/j.trc.2023.104062} {10.1016/j.trc.2023.104062}}.

\bibitem{wang2022modeling}
B.~Wang, Z.~Li, S.~Wang, M.~Li, and A.~Ji, ``Modeling bounded rationality in discretionary lane change with the quantal response equilibrium of game theory,'' \emph{Transp. Res. Part B, Methodol.}, vol. 164, pp. 145--161, 2022, doi: {\color{blue} \href{https://doi.org/10.1016/j.trb.2022.08.008} {10.1016/j.trb.2022.08.008}}.

\bibitem{karimi2023level}
S.~Karimi, A.~Karimi, and A.~Vahidi, ``Level-$k$ reasoning, deep reinforcement learning, and monte carlo decision process for fast and safe automated lane change and speed management,'' \emph{IEEE Trans. Intell. Veh.}, vol.~8, no.~6, pp. 3556--3571, 2023, doi: {\color{blue} \href{ https://doi.org/10.1109/TIV.2023.3265311} {10.1109/TIV.2023.3265311}}.

\bibitem{Sun2022micro}
Z.~Sun, Z.~Qin, R.~Ma, T.~Huang, Z.~Gao, and A.~Ji, ``Microscopic right-of-way trading mechanism for cooperative decision-making: Theories and preliminary results,'' \emph{SSRN Electron. J.}, 2023, doi: {\color{blue} \href{http://dx.doi.org/10.2139/ssrn.4571173} {10.2139/ssrn.4571173}}.

\bibitem{liao2021game}
X.~Liao, X.~Zhao, Z.~Wang, K.~Han, P.~Tiwari, M.~J. Barth, and G.~Wu, ``Game theory-based ramp merging for mixed traffic with unity-sumo co-simulation,'' \emph{IEEE Trans. Syst. Man Cybern.: Syst.}, vol.~52, no.~9, pp. 5746--5757, 2021, doi: {\color{blue} \href{http://dx.doi.org/10.1109/TSMC.2021.3131431} {10.1109/TSMC.2021.3131431}}.

\bibitem{li2023simulation}
W.~Li, F.~Qiu, L.~Li, Y.~Zhang, and K.~Wang, ``Simulation of vehicle interaction behavior in merging scenarios: A deep maximum entropy- inverse reinforcement learning method combined with game theory,'' \emph{IEEE Trans. Intell. Veh.}, pp. 1--15, 2023, doi: {\color{blue} \href{ https://doi.org/10.1109/TIV.2023.3323138} {10.1109/TIV.2023.3323138}}.

\bibitem{sun2020microscopic}
Z.~Sun, T.~Huang, Z.~Qin, and Z.~Gao, ``Microscopic right-of-way trading mechanism for cooperative decision-making: A ramp merging example,'' in \emph{Proc. Transp. Res. Board 99th Annu. Meeting (TRB), Washington, DC}, 2020.

\bibitem{sayin2018information}
M.~O. Sayin, C.-W. Lin, S.~Shiraishi, J.~Shen, and T.~Ba{\c{s}}ar, ``Information-driven autonomous intersection control via incentive compatible mechanisms,'' \emph{{IEEE} Trans. Intell. Transp. Syst.}, vol.~20, no.~3, pp. 912--924, 2018, doi: {\color{blue} \href{http://dx.doi.org/10.1109/TITS.2018.2838049} {10.1109/TITS.2018.2838049}}.

\bibitem{wang2019enabling}
Y.~Wang, Y.~Ren, S.~Elliott, and W.~Zhang, ``Enabling courteous vehicle interactions through game-based and dynamics-aware intent inference,'' \emph{IEEE Trans. Intell. Veh.}, vol.~5, no.~2, pp. 217--228, 2020, doi: {\color{blue} \href{ https://doi.org/10.1109/TIV.2019.2955897} {10.1109/TIV.2019.2955897}}.

\bibitem{wang2021game}
K.~Wang, Y.~Wang, H.~Du, and K.~Nam, ``Game-theory-inspired hierarchical distributed control strategy for cooperative intersection considering priority negotiation,'' \emph{{IEEE} Trans. Veh. Technol.}, vol.~70, no.~7, pp. 6438--6449, 2021, doi: {\color{blue} \href{http://dx.doi.org/10.1109/TVT.2021.3086563} {10.1109/TVT.2021.3086563}}.

\bibitem{lloret2016envy}
R.~Lloret-Batlle and R.~Jayakrishnan, ``Envy-minimizing pareto efficient intersection control with brokered utility exchanges under user heterogeneity,'' \emph{Transp. Res. Part B, Methodol.}, vol.~94, pp. 22--42, 2016, doi: {\color{blue} \href{https://doi.org/10.1016/j.trb.2016.08.014} {10.1016/j.trb.2016.08.014}}.

\bibitem{abdelghaffar2019development}
H.~M. Abdelghaffar and H.~A. Rakha, ``Development and testing of a novel game theoretic de-centralized traffic signal controller,'' \emph{{IEEE} Trans. Intell. Transp. Syst.}, vol.~22, no.~1, pp. 231--242, 2019, doi: {\color{blue} \href{https://doi.org/10.1109/TITS.2019.2955918} {10.1109/TITS.2019.2955918}}.

\bibitem{sun2021modeling}
Z.~Sun, X.~Yao, Z.~Qin, P.~Zhang, and Z.~Yang, ``Modeling car-following heterogeneities by considering leader--follower compositions and driving style differences,'' \emph{Transp. Res. Rec.}, vol. 2675, no.~11, pp. 851--864, 2021, doi: {\color{blue} \href{ https://doi.org/10.1177/03611981211020006} {10.1177/03611981211020006}}.

\bibitem{zhang2014analysis}
L.~Zhang, X.~W. huang, and W.~M. Wu, ``The analysis of driver's behavior in non-signalized intersection based on the game,'' \emph{Appl. Mech. Mater.}, vol. 505, pp. 1157--1162, 2014, doi: {\color{blue} \href{https://doi.org/10.4028/www.scientific.net/AMM.505-506.1157} {10.4028/www.scientific.net/AMM.505-506.1157}}.

\bibitem{zhang2021trajectory}
Y.~Zhang, R.~Hao, T.~Zhang, X.~Chang, Z.~Xie, and Q.~Zhang, ``A trajectory optimization-based intersection coordination framework for cooperative autonomous vehicles,'' \emph{{IEEE} Trans. Intell. Transp. Syst.}, 2021, doi: {\color{blue} \href{http://dx.doi.org/10.1109/TITS.2021.3131570} {10.1109/TITS.2021.3131570}}.

\bibitem{wolshon2016traffic}
B.~Wolshon, A.~Pande \emph{et~al.}, \emph{Traffic engineering handbook}.\hskip 1em plus 0.5em minus 0.4em\relax John Wiley \& Sons, 2016.

\bibitem{zhu2015linear}
F.~Zhu and S.~V. Ukkusuri, ``A linear programming formulation for autonomous intersection control within a dynamic traffic assignment and connected vehicle environment,'' \emph{Transp. Res. Part C, Emerg. Technol.}, vol.~55, pp. 363--378, 2015, doi: {\color{blue} \href{https://doi.org/10.1016/j.trc.2015.01.006} {10.1016/j.trc.2015.01.006}}.

\bibitem{hancock2013policy}
M.~W. Hancock and B.~Wright, ``A policy on geometric design of highways and streets,'' \emph{American Association of State Highway and Transportation Officials: Washington, DC, USA}, vol.~3, 2013.

\bibitem{rodegerdts2004signalized}
L.~A. Rodegerdts, B.~L. Nevers, B.~Robinson, J.~Ringert, P.~Koonce, J.~Bansen, T.~Nguyen, J.~McGill, D.~Stewart, J.~Suggett \emph{et~al.}, ``Signalized intersections: informational guide,'' United States. Federal Highway Administration, Tech. Rep., 2004.

\bibitem{Bui2017}
K.~H.~N. Bui, J.~E. Jung, and D.~Camacho, ``{Game theoretic approach on Real-time decision making for IoT-based traffic light control},'' \emph{Concurrency Comput. Pract. Exper.}, vol.~29, no.~11, 2017, doi: {\color{blue} \href{http://dx.doi.org/10.1002/cpe.4077} {10.1002/cpe.4077}}.

\bibitem{Zhao2019}
Y.~Zhao, Y.~Liang, J.~Hu, and Z.~Zhang, ``{Traffic Signal Control for Isolated Intersection Based on Coordination Game and Pareto Efficiency},'' in \emph{Proc. 2019 IEEE Intell. Transp. Syst. Conf.}, Auckland, New Zealand, 2019, pp. 3508--3513, doi: {\color{blue} \href{http://dx.doi.org/10.1109/ITSC.2019.8917165} {10.1109/ITSC.2019.8917165}}.

\bibitem{Khanjary2013}
M.~Khanjary, ``Using game theory to optimize traffic light of an intersection,'' in \emph{Proc. 14th Int. Symp. Comput. Intell. Inform.}, Budapest, Hungary, 2013, pp. 249--253, doi: {\color{blue}\href{http://dx.doi.org/10.1109/CINTI.2013.6705201} {10.1109/CINTI.2013.6705201}}.

\bibitem{li2018game}
N.~Li, I.~Kolmanovsky, A.~Girard, and Y.~Yildiz, ``Game theoretic modeling of vehicle interactions at unsignalized intersections and application to autonomous vehicle control,'' in \emph{Proc. Annu. Amer. Control Conf. (ACC)}, 2018, pp. 3215--3220, doi: {\color{blue} \href{http://dx.doi.org/10.23919/ACC.2018.8430842} {10.23919/ACC.2018.8430842}}.

\bibitem{elhenawy2015intersection}
M.~Elhenawy, A.~A. Elbery, A.~A. Hassan, and H.~A. Rakha, ``An intersection game-theory-based traffic control algorithm in a connected vehicle environment,'' in \emph{Proc. IEEE 18th Int. Conf. Intell. Transp. Syst}, Gran Canaria, Spain, 2015, pp. 343--347, doi: {\color{blue} \href{http://dx.doi.org/10.1109/ITSC.2015.65} {10.1109/ITSC.2015.65}}.

\bibitem{cheng2018speed}
C.~Cheng, Z.~Yang, and D.~Yao, ``A speed guide model for collision avoidance in non-signalized intersections based on reduplicate game theory,'' in \emph{Proc. IEEE Intell. Vehicles Symp. (IV)}, Changshu, China, 2018, pp. 1614--1619, doi: {\color{blue} \href{http://dx.doi.org/10.1109/IVS.2018.8500502} {10.1109/IVS.2018.8500502}}.

\bibitem{chen2020conflict}
X.~Chen, Y.~Sun, Y.~Ou, X.~Zheng, Z.~Wang, and M.~Li, ``A conflict decision model based on game theory for intelligent vehicles at urban unsignalized intersections,'' \emph{IEEE Access}, vol.~8, pp. 189\,546--189\,555, 2020, doi: {\color{blue} \href{http://dx.doi.org/10.1109/ACCESS.2020.3031674} {10.1109/ACCESS.2020.3031674}}.

\bibitem{jin2020game}
X.~Jin, K.~Li, Q.-S. Jia, H.~Xia, Y.~Bai, and D.~Ren, ``A game-theoretic reinforcement learning approach for adaptive interaction at intersections,'' in \emph{Proc. Chin. Autom. Congr. (CAC)}, Shanghai, China, 2020, pp. 4451--4456, doi: {\color{blue} \href{http://dx.doi.org/10.1109/CAC51589.2020.9327245} {10.1109/CAC51589.2020.9327245}}.

\bibitem{tian2018adaptive}
R.~Tian, S.~Li, N.~Li, I.~Kolmanovsky, A.~Girard, and Y.~Yildiz, ``Adaptive game-theoretic decision making for autonomous vehicle control at roundabouts,'' in \emph{Proc. IEEE Conf. Decis. Control (CDC)}, Miami, FL, USA, Dec. 2018, pp. 321--326, doi: {\color{blue} \href{http://dx.doi.org/10.1109/CDC.2018.8619275} {10.1109/CDC.2018.8619275}}.

\bibitem{zohdy2012game}
I.~H. Zohdy and H.~Rakha, ``Game theory algorithm for intersection-based cooperative adaptive cruise control (cacc) systems,'' in \emph{Proc. IEEE 15th Int. Conf. Intell. Transp. Syst.}, Anchorage, AK, USA, Sep. 2012, pp. 1097--1102, doi: {\color{blue} \href{http://dx.doi.org/10.1109/ITSC.2012.6338644} {10.1109/ITSC.2012.6338644}}.

\bibitem{yang2016cooperative}
Z.~Yang, H.~Huang, D.~Yao, and Y.~Zhang, ``Cooperative driving model for non-signalized intersections based on reduplicate dynamic game,'' in \emph{Proc. IEEE 19th Int. Conf. Intell. Transp. Syst. (ITSC)}, Rio de Janeiro, Nov. 2016, pp. 1366--1371, doi: {\color{blue} \href{http://dx.doi.org/10.1109/ITSC.2016.7795735} {10.1109/ITSC.2016.7795735}}.

\bibitem{fan2014characteristics}
H.~Fan, B.~Jia, J.~Tian, and L.~Yun, ``Characteristics of traffic flow at a non-signalized intersection in the framework of game theory,'' \emph{Phys. A, Statist. Mech. Appl.}, vol. 415, pp. 172--180, 2014, doi: {\color{blue} \href{https://doi.org/10.1016/j.physa.2014.07.031} {10.1016/j.physa.2014.07.031}}.

\bibitem{zhang2012dilemma}
W.~Zhang and W.~Chen, ``Dilemma game in a cellular automaton model with a non-signalized intersection,'' \emph{Eur. Phys. J. B}, vol.~85, no.~2, pp. 1--8, 2012, doi: {\color{blue} \href{https://doi.org/10.1140/epjb/e2012-20904-x} {10.1140/epjb/e2012-20904-x}}.

\bibitem{wang2011dirty}
L.~Wang, N.~G. Xie, and R.~Meng, ``Dirty-face game analysis on mixed traffic flow at unsignalized intersection,'' \emph{Adv. Mat. Res.}, vol. 201, pp. 2119--2125, 2011, doi: {\color{blue} \href{https://doi.org/10.4028/www.scientific.net/AMR.201-203.2119} {10.4028/www.scientific.net/AMR.201-203.2119}}.

\bibitem{lemmer2020driver}
M.~Lemmer, S.~Schwab, and S.~Hohmann, ``Driver interaction at intersections: A hybrid dynamic game based model,'' in \emph{Proc. IEEE Int. Conf. Syst., Man, Cybern. (SMC)}, Toronto, ON, Canada, Oct. 2020, pp. 2269--2276, doi: {\color{blue} \href{http://dx.doi.org/10.1109/SMC42975.2020.9283079} {10.1109/SMC42975.2020.9283079}}.

\bibitem{hang2022driving}
P.~Hang, C.~Huang, Z.~Hu, and C.~Lv, ``Driving conflict resolution of autonomous vehicles at unsignalized intersections: A differential game approach,'' \emph{IEEE/ASME Trans. Mechatron.}, 2022, doi: {\color{blue} \href{http://dx.doi.org/10.1109/TMECH.2022.3174273} {10.1109/TMECH.2022.3174273}}.

\bibitem{qiu2011non}
X.~Qiu, L.~Yang, L.~Zhang, and Z.~Huang, ``The non-signal intersection traffic behavior analysis based on game theory,'' in \emph{Proc. 4th Int. Symp. Comput. Intell. Design (ISCID)}, vol.~2, Hangzhou, China, Oct. 2011, pp. 122--124, doi: {\color{blue} \href{http://dx.doi.org/10.1109/ISCID.2011.132} {10.1109/ISCID.2011.132}}.

\bibitem{bouderba2019evolutionary}
S.~I. Bouderba and N.~Moussa, ``Evolutionary dilemma game for conflict resolution at unsignalized traffic intersection,'' \emph{Int. J. Mod. Phys. C}, vol.~30, no. 02n03, p. 1950018, 2019, doi: {\color{blue} \href{https://doi.org/10.1142/S0129183119500189} {10.1142/S0129183119500189}}.

\bibitem{cai2021game}
J.~Cai, P.~Hang, and C.~Lv, ``Game theoretic modeling and decision making for connected vehicle interactions at urban intersections,'' in \emph{Proc. IEEE Int. Conf. Adv. Robot. Mech (ICARM)}, Chongqing, China, Jul. 2021, pp. 874--880, doi: {\color{blue} \href{http://dx.doi.org/10.1109/ICARM52023.2021.9536147} {10.1109/ICARM52023.2021.9536147}}.

\bibitem{qi2014game}
W.~Qi, H.~Wen, C.~Fu, and M.~Song, ``Game theory model of traffic participants within amber time at signalized intersection,'' \emph{Comput. Intell. Neurosci.}, vol. 2014, pp. 56--56, 2014, doi: {\color{blue} \href{http://dx.doi.org/10.1155/2014/756235} {10.1155/2014/756235}}.

\bibitem{liu2020game}
G.~Liu, B.~Xiao, and D.~Li, ``Game-theory based driving decision algorithm for intersection scenarios considering driver irrationality,'' in \emph{Proc. IEEE 4th CAA Int. Conf. Veh. Control Intell (CVCI)}, Hangzhou, China, Dec. 2020, pp. 747--752, doi: {\color{blue}\href{http://dx.doi.org/10.1109/CVCI51460.2020.9338515} {10.1109/CVCI51460.2020.9338515}}.

\bibitem{li2022human}
D.~Li, G.~Liu, and B.~Xiao, ``Human-like driving decision at unsignalized intersections based on game theory,'' \emph{Proc. Inst. Mech. Eng., Part D: J. Automob. Eng.}, pp. 159--173, 2022, doi: {\color{blue}\href{http://dx.doi.org/10.1177/09544070221075423} {10.1177/09544070221075423}}.

\bibitem{baz2020intersection}
A.~Baz, P.~Yi, and A.~Qurashi, ``Intersection control and delay optimization for autonomous vehicles flows only as well as mixed flows with ordinary vehicles,'' \emph{Vehicles}, vol.~2, no.~3, pp. 523--541, 2020, doi: {\color{blue}\href{http://dx.doi.org/10.3390/vehicles2030029} {10.3390/vehicles2030029}}.

\bibitem{2018Intersection}
H.~Wei, L.~Mashayekhy, and J.~Papineau, ``Intersection management for connected autonomous vehicles: A game theoretic framework,'' in \emph{Proc. 21st Int. Conf. Intell. Transp. Syst. (ITSC)}, Maui, HI, USA, Nov. 2018, doi: {\color{blue}\href{http://dx.doi.org/10.1109/ITSC.2018.8569307} {10.1109/ITSC.2018.8569307}}.

\bibitem{Chen2017Modeling}
M.~Liu, Y.~Chen, G.~Lu, and Y.~Wang, ``Modeling crossing behavior of drivers at unsignalized intersections with consideration of risk perception,'' \emph{Transp. Res. Part F, Traffic Psychol. Behav.}, vol.~45, pp. 14--26, 2017, doi: {\color{blue}\href{https://doi.org/10.1016/j.trf.2016.11.012} {10.1016/j.trf.2016.11.012}}.

\bibitem{2020Driver}
M.~Lemmer, S.~Schwab, and S.~Hohmann, ``Driver interaction at intersections: A hybrid dynamic game based model,'' in \emph{Proc. IEEE Int. Conf. Syst., Man, Cybern. (SMC)}, Toronto, ON, Canada, Oct. 2020, doi: {\color{blue}\href{https://doi.org/10.1109/SMC42975.2020.9283079} {10.1109/SMC42975.2020.9283079}}.

\bibitem{lin2020comparative}
D.~Lin and S.~E. Jabari, ``Comparative analysis of economic instruments in intersection operation: A user-based perspective,'' in \emph{Proc. 2020 IEEE 23rd Int. Conf. Int. Transp. Syst. (ITSC)}, Rhodes, Greece, Sep. 2020, pp. 1--6, doi: {\color{blue}\href{https://doi.org/10.1109/ITSC45102.2020.9294641} {10.1109/ITSC45102.2020.9294641}}.

\bibitem{2018Reliable}
M.~O. Sayin, C.-W. Lin, S.~Shiraishi, and T.~Ba{\c{s}}ar, ``Reliable intersection control in non-cooperative environments,'' in \emph{Proc. Annu. Amer. Control Conf. (ACC)}, Milwaukee, WI, USA, Jun. 2018, pp. 609--614, doi: {\color{blue}\href{https://doi.org/10.23919/ACC.2018.8430809} {10.23919/ACC.2018.8430809}}.

\bibitem{Li2021safe}
D.~Li, A.~Liu, H.~Pan, and W.~Chen, ``Safe, efficient and socially-compatible decision of automated vehicles: a case study of unsignalized intersection driving,'' \emph{Automot. Innov.}, pp. 1--16, 2023, doi: {\color{blue}\href{https://doi.org/10.1007/s42154-023-00219-2} {10.1007/s42154-023-00219-2}}.

\bibitem{liu2022three}
M.~Liu, Y.~Wan, F.~L. Lewis, S.~Nageshrao, and D.~Filev, ``A three-level game-theoretic decision-making framework for autonomous vehicles,'' \emph{{IEEE} Trans. Intell. Transp. Syst.}, vol.~23, no.~11, pp. 20\,298--20\,308, 2022, doi: {\color{blue}\href{https://doi.org/10.1109/TITS.2022.3172926} {10.1109/TITS.2022.3172926}}.

\bibitem{Zhu2022}
D.~Zhu, N.~N. Sze, Z.~Feng, and Z.~Yang, ``{A two-stage safety evaluation model for the red light running behaviour of pedestrians using the game theory},'' \emph{Safety Sci.}, vol. 147, no. November 2021, p. 105600, 2022, doi: {\color{blue}\href{https://doi.org/10.1016/j.ssci.2021.105600} {10.1016/j.ssci.2021.105600}}.

\bibitem{Yang2020}
C.~Yang, J.~Wang, and J.~Dong, ``{Capacity Model of Exclusive Right-Turn Lane at Signalized Intersection considering Pedestrian-Vehicle Interaction},'' \emph{J. Adv. Transp.}, vol. 2020, 2020, doi: {\color{blue}\href{https://doi.org/10.1155/2020/1534564} {10.1155/2020/1534564}}.

\bibitem{Wang2021}
H.~Wang, Q.~Meng, S.~Chen, and X.~Zhang, ``{Competitive and cooperative behaviour analysis of connected and autonomous vehicles across unsignalised intersections: A game-theoretic approach},'' \emph{Transp. Res. Part B, Methodol.}, vol. 149, pp. 322--346, 2021, doi: {\color{blue}\href{https://doi.org/10.1016/j.trb.2021.05.007} {10.1016/j.trb.2021.05.007}}.

\bibitem{Rahmati2021}
Y.~Rahmati, M.~K. Hosseini, and A.~Talebpour, ``{Helping Automated Vehicles With Left-Turn Maneuvers: A Game Theory-Based Decision Framework for Conflicting Maneuvers at Intersections},'' \emph{{IEEE} Trans. Intell. Transp. Syst.}, pp. 1--14, 2021, doi: {\color{blue}\href{https://doi.org/10.1109/TITS.2021.3108409} {10.1109/TITS.2021.3108409}}.

\bibitem{Arbis2016}
D.~Arbis, V.~V. Dixit, and T.~H. Rashidi, ``{Impact of risk attitudes and perception on game theoretic driving interactions and safety},'' \emph{Accid. Anal. Prev.}, vol.~94, pp. 135--142, 2016, doi: {\color{blue}\href{https://doi.org/10.1016/j.aap.2016.05.027} {10.1016/j.aap.2016.05.027}}.

\bibitem{lu2023game}
X.~Lu, H.~Zhao, C.~Li, B.~Gao, and H.~Chen, ``A game-theoretic approach on conflict resolution of autonomous vehicles at unsignalized intersections,'' \emph{{IEEE} Trans. Intell. Transp. Syst.}, 2023, doi: {\color{blue} \href{http://doi.org/10.1109/TITS.2023.3285597} {10.1109/TITS.2023.3285597}}.

\bibitem{jia2023interactive}
S.~Jia, Y.~Zhang, X.~Li, X.~Na, Y.~Wang, B.~Gao, B.~Zhu, and R.~Yu, ``Interactive decision-making with switchable game modes for automated vehicles at intersections,'' \emph{{IEEE} Trans. Intell. Transp. Syst.}, 2023, doi: {\color{blue} \href{http://doi.org/10.1109/TITS.2023.3286075} {10.1109/TITS.2023.3286075}}.

\bibitem{Molinari2018}
F.~Molinari and J.~Raisch, ``{Automation of Road Intersections Using Consensus-based Auction Algorithms},'' in \emph{Proc. Annu. Amer. Control Conf. (ACC)}, vol. 2018-June, Milwaukee, WI, USA, Jun. 2018, pp. 5994--6001, doi: {\color{blue}\href{https://doi.org/10.23919/ACC.2018.8430865} {10.23919/ACC.2018.8430865}}.

\bibitem{Xu2019}
Y.~Xu, D.~Li, and Y.~Xi, ``{A Game-Based Adaptive Traffic Signal Control Policy Using the Vehicle to Infrastructure (V2I)},'' \emph{{IEEE} Trans. Veh. Technol.}, vol.~68, no.~10, pp. 9425--9437, 2019, doi: {\color{blue}\href{https://doi.org/10.1109/TVT.2019.2933317} {10.1109/TVT.2019.2933317}}.

\bibitem{Vasirani2012}
M.~Vasirani and S.~Ossowski, ``{A Market-Inspired Approach for Intersection Management in Urban Road Traffic Networks},'' \emph{J. Artif. Intell. Res.}, vol.~43, pp. 621--659, 2012, doi: {\color{blue}\href{https://doi.org/10.1613/jair.3560} {10.1613/jair.3560}}.

\bibitem{Cheng2019}
C.~Cheng, D.~Yao, Y.~Zhang, J.~Li, and Y.~Guo, ``{A Vehicle Passing Model in Non-signalized Intersections based on Non-cooperative Game Theory},'' \emph{Proc. 2019 IEEE Intell. Transp. Syst. Conf.}, pp. 2286--2291, Oct. 2019, doi: {\color{blue}\href{https://doi.org/10.1109/ITSC.2019.8917396} {10.1109/ITSC.2019.8917396}}.

\bibitem{Cabri2021}
G.~Cabri, L.~Gherardini, M.~Montangero, and F.~Muzzini, ``{About auction strategies for intersection management when human-driven and autonomous vehicles coexist},'' \emph{Multimed. Tools. Appl.}, pp. 15\,921--15\,936, 2021, doi: {\color{blue} \href{http://dx.doi.org/10.1007/s11042-020-10222-y} {10.1007/s11042-020-10222-y}}.

\bibitem{Carlino2013}
D.~Carlino, S.~D. Boyles, and P.~Stone, ``Auction-based autonomous intersection management,'' in \emph{Proc. 16th Int. IEEE Conf. Intell. Transp. Syst. (ITSC)}, The Hague, Netherlands, Oct. 2013, pp. 529--534, doi: {\color{blue} \href{http://dx.doi.org/10.1109/ITSC.2013.6728285} {10.1109/ITSC.2013.6728285}}.

\bibitem{Schepperle2008}
H.~Schepperle and K.~B{\"{o}}hm, ``{Auction-based traffic management: Towards effective concurrent utilization of road intersections},'' in \emph{Proc. 2008 10th IEEE Conf. E-Commerce Tech. 5th IEEE Conf. Enterprise Comput., E-Commerce E-Services}, Arlington, VA, USA, 2008, pp. 105--112, doi: {\color{blue} \href{http://dx.doi.org/10.1109/CECandEEE.2008.88} {10.1109/CECandEEE.2008.88}}.

\bibitem{hang2022decision}
P.~Hang, C.~Huang, Z.~Hu, and C.~Lv, ``Decision making for connected automated vehicles at urban intersections considering social and individual benefits,'' \emph{{IEEE} Trans. Intell. Transp. Syst.}, vol.~23, no.~11, pp. 22\,549--22\,562, 2022, doi: {\color{blue} \href{http://dx.doi.org/10.1109/TITS.2022.3209607} {10.1109/TITS.2022.3209607}}.

\bibitem{suriyarachchi2022gameopt}
N.~Suriyarachchi, R.~Chandra, J.~S. Baras, and D.~Manocha, ``Gameopt: Optimal real-time multi-agent planning and control for dynamic intersections,'' in \emph{Proc. IEEE 25th Int. Conf. Intell. Transp. Syst.}, Macau, China, Nov. 2022, pp. 2599--2606, doi: {\color{blue} \href{http://dx.doi.org/10.1109/ITSC55140.2022.9921968} {10.1109/ITSC55140.2022.9921968}}.

\bibitem{chandra2022gameplan}
R.~Chandra and D.~Manocha, ``{GamePlan: Game-Theoretic Multi-Agent Planning with Human Drivers at Intersections, Roundabouts, and Merging},'' \emph{{IEEE} Robot. Autom. Lett.}, vol.~7, no.~2, pp. 2676--2683, 2022, doi: {\color{blue} \href{http://dx.doi.org/10.1109/LRA.2022.3144516} {10.1109/LRA.2022.3144516}}.

\bibitem{Molinari2019}
F.~Molinari, A.~M. Dethof, and J.~Raisch, ``{Traffic automation in urban road networks using consensus-based auction algorithms for road intersections},'' \emph{Proc. 18th Eur. Control Conf.}, pp. 3008--3015, Jun. 2019, doi: {\color{blue} \href{http://dx.doi.org/10.23919/ECC.2019.8796170} {10.23919/ECC.2019.8796170}}.

\bibitem{Tian2022}
R.~Tian, N.~Li, I.~Kolmanovsky, Y.~Yildiz, and A.~R. Girard, ``{Game-Theoretic Modeling of Traffic in Unsignalized Intersection Network for Autonomous Vehicle Control Verification and Validation},'' \emph{{IEEE} Trans. Intell. Transp. Syst.}, vol.~23, no.~3, pp. 2211--2226, 2022, doi: {\color{blue} \href{http://dx.doi.org/10.1109/TITS.2020.3035363} {10.1109/TITS.2020.3035363}}.

\bibitem{Levin2015}
M.~W. Levin and S.~D. Boyles, ``{Intersection auctions and reservation-based control in dynamic traffic assignment},'' \emph{Transp. Res. Rec.}, vol. 2497, pp. 35--44, 2015, doi: {\color{blue} \href{http://dx.doi.org/10.3141/2497-04} {10.3141/2497-04}}.

\bibitem{Liu2020}
M.~Liu, S.~Zhang, and L.~Teng, ``{Method of Controlling Vehicles at Intersections Based on the Auction Algorithm},'' \emph{J. Phys. Conf. Ser.}, vol. 1453, no.~1, 2020, doi: {\color{blue} \href{http://dx.doi.org/10.1088/1742-6596/1453/1/012050} {10.1088/1742-6596/1453/1/012050}}.

\bibitem{Wang2020}
H.~Wang, Y.~Li, and H.~V. Zhao, ``{Performance analysis of road intersections based on game theory and dynamic level-k model},'' \emph{Proc. IEEE Int. Conf Parallel Distrib. Process. With Appl., Big Data Cloud Comput., Sustain. Comput. Commun., Social Comput. Netw. (ISPA/BDCloud/SocialCom/SustainCom)}, pp. 1112--1119, Dec. 2020, doi: {\color{blue} \href{http://dx.doi.org/10.1109/ISPA-BDCloud-SocialCom-SustainCom51426.2020.00166} {10.1109/ISPA-BDCloud-SocialCom-SustainCom51426.2020.00166}}.

\bibitem{Philippe2019}
C.~Philippe, L.~Adouane, A.~Tsourdos, H.-S. Shin, and B.~Thuilot, ``Probability collectives algorithm applied to decentralized intersection coordination for connected autonomous vehicles,'' in \emph{Proc. IEEE Intell. Vehicles Symp. (IV)}, Paris, France, Jun. 2019, pp. 1928--1934, doi: {\color{blue} \href{http://dx.doi.org/10.1109/IVS.2019.8813827} {10.1109/IVS.2019.8813827}}.

\bibitem{Schwarting2019}
W.~Schwarting, A.~Pierson, J.~Alonso-Mora, S.~Karaman, and D.~Rus, ``{Social behavior for autonomous vehicles},'' \emph{Proc. Natl. Acad. Sci.}, vol. 116, no.~50, pp. 2492--24\,978, 2019, doi: {\color{blue} \href{http://dx.doi.org/10.1073/pnas.1820676116} {10.1073/pnas.1820676116}}.

\bibitem{Sarkar2021}
A.~Sarkar and K.~Czarnecki, ``{Solution Concepts in Hierarchical Games Under Bounded Rationality With Applications to Autonomous Driving},'' in \emph{Proc. 35th Conf. Artif. Intell. (AAAI)}, vol.~6B, no. Samuelson 1995, 2021, pp. 5698--5708, doi: {\color{blue} \href{http://dx.doi.org/10.1609/aaai.v35i6.16715} {10.1609/aaai.v35i6.16715}}.

\bibitem{yuan2021deep}
M.~Yuan, J.~Shan, and K.~Mi, ``Deep reinforcement learning based game-theoretic decision-making for autonomous vehicles,'' \emph{{IEEE} Robot. Autom. Lett.}, vol.~7, no.~2, pp. 818--825, 2021, doi: {\color{blue} \href{http://dx.doi.org/10.1109/LRA.2021.3134249} {10.1109/LRA.2021.3134249}}.

\bibitem{Rey2021}
D.~Rey, M.~W. Levin, and V.~V. Dixit, ``{Online incentive-compatible mechanisms for traffic intersection auctions},'' \emph{Eur. J. Oper. Res.}, vol. 293, no.~1, pp. 229--247, 2021, doi: {\color{blue} \href{http://dx.doi.org/10.1016/j.ejor.2020.12.030} {10.1016/j.ejor.2020.12.030}}.

\bibitem{Tan2010}
L.~Tan, X.~Zhao, D.~Hu, Y.~Shang, and W.~Ren, ``A study of single intersection traffic signal control based on two-player cooperation game model,'' in \emph{Proc. Int. Conf. Inf. Eng}, vol.~2, Beidai, China, Aug. 2010, pp. 322--327, doi: {\color{blue} \href{http://dx.doi.org/10.1109/ICIE.2010.172} {10.1109/ICIE.2010.172}}.

\bibitem{tan2017multi}
L.~Tan, Y.~Wang, Q.~Wang, Y.~Sun, and Z.~Li, ``Multi-phase signal control method for single intersection based on multi-person cooperative game theory,'' in \emph{Proc. 2017 2nd Jt. Int. Inf. Technol. Mech. Electron. Eng. Conf. (JIMEC 2017)}, Oct. 2017, pp. 335--339, doi: {\color{blue} \href{http://dx.doi.org/10.2991/jimec-17.2017.74} {10.2991/jimec-17.2017.74}}.

\bibitem{NamBui2018}
K.~H. {Nam Bui} and J.~J. Jung, ``{Cooperative game-theoretic approach to traffic flow optimization for multiple intersections},'' \emph{Comput. Electr. Eng.}, vol.~71, pp. 1012--1024, 2018, doi: {\color{blue}\href{http://dx.doi.org/10.1016/j.compeleceng.2017.10.016} {10.1016/j.compeleceng.2017.10.016}}.

\bibitem{Dong2011}
H.~Dong and Z.~Dai, ``{A multi intersections signal coordinate control method based on game theory},'' \emph{Proc. Int. Conf. Electron., Commun. Control (ICECC)}, no.~3, pp. 1232--1235, 2011, doi: {\color{blue}\href{http://dx.doi.org/10.1109/ICECC.2011.6066604} {10.1109/ICECC.2011.6066604}}.

\bibitem{Raphael2017}
J.~Raphael, E.~I. Sklar, and S.~Maskell, ``An intersection-centric auction-based traffic signal control framework,'' in \emph{Agent-Based Modeling of Sustainable Behaviors}.\hskip 1em plus 0.5em minus 0.4em\relax Cham, Switzerland: Springer International Publishing, 2017, pp. 121--142, doi: {\color{blue}\href{http://dx.doi.org/10.1007/978-3-319-46331-5\_6} {10.1007/978-3-319-46331-5\_6}}.

\bibitem{Xia2018Adaptive}
X.-h. Xia, ``Adaptive traffic signal coordinated timing decision for adjacent intersections with chicken game,'' in \emph{Intelligent Transport Systems -- From Research and Development to the Market Uptake}, T.~Kov{\'a}{\v{c}}ikov{\'a}, {\v{L}}.~Buzna, G.~Pourhashem, G.~Lugano, Y.~Cornet, and N.~Lugano, Eds.\hskip 1em plus 0.5em minus 0.4em\relax Cham: Springer International Publishing, 2018, pp. 239--251, doi: {\color{blue}\href{http://dx.doi.org/10.1007/978-3-319-93710-6\_25} {10.1007/978-3-319-93710-6\_25}}.

\bibitem{Abdoos2021}
M.~Abdoos, ``A cooperative multiagent system for traffic signal control using game theory and reinforcement learning,'' \emph{{IEEE} Intell. Transp. Syst. Mag.}, vol.~13, no.~4, pp. 6--16, 2021, doi: {\color{blue}\href{http://dx.doi.org/10.1109/MITS.2020.2990189} {10.1109/MITS.2020.2990189}}.

\bibitem{Clempner2015}
J.~B. Clempner and A.~S. Poznyak, ``{Modeling the multi-traffic signal-control synchronization: A Markov chains game theory approach},'' \emph{Eng. Appl. Artif. Intell.}, vol.~43, pp. 147--156, 2015, doi: {\color{blue}\href{http://dx.doi.org/10.1016/j.engappai.2015.04.009} {10.1016/j.engappai.2015.04.009}}.

\bibitem{CastilloGonzalez2019}
R.~{Castillo Gonz{\'{a}}lez}, J.~B. Clempner, and A.~S. Poznyak, ``{Solving traffic queues at controlled-signalized intersections in continuous-time Markov games},'' \emph{Math. Comput. Simul.}, vol. 166, pp. 283--297, 2019, doi: {\color{blue}\href{http://dx.doi.org/10.1016/j.matcom.2019.06.002} {10.1016/j.matcom.2019.06.002}}.

\bibitem{Xu2013}
L.~H. Xu, X.~H. Xia, and Q.~Luo, ``{The study of reinforcement learning for traffic self-adaptive control under multiagent Markov game environment},'' \emph{Math. Probl. Eng.}, vol. 2013, 2013, doi: {\color{blue}\href{http://dx.doi.org/10.1155/2013/962869} {10.1155/2013/962869}}.

\bibitem{ZhuYT2022}
Y.~Zhu, Z.~He, and G.~Li, ``A bi-hierarchical game-theoretic approach for network-wide traffic signal control using trip-based data,'' \emph{{IEEE} Trans. Intell. Transp. Syst.}, vol.~23, no.~9, pp. 15\,408--15\,419, 2022, doi: {\color{blue}\href{http://dx.doi.org/10.1109/TITS.2022.3140511} {10.1109/TITS.2022.3140511}}.

\bibitem{Stryszowski2021}
M.~Stryszowski, S.~Longo, D.~D'Alessandro, E.~Velenis, G.~Forostovsky, and S.~Manfredi, ``{A Framework for Self-Enforced Optimal Interaction between Connected Vehicles},'' \emph{{IEEE} Trans. Intell. Transp. Syst.}, vol.~22, no.~10, pp. 6152--6161, 2021, doi: {\color{blue}\href{https://doi.org/10.1109/TITS.2020.2988150} {10.1109/TITS.2020.2988150}}.

\bibitem{Dai2013}
Z.~Dai, H.~Dong, and Q.~Wang, ``A multi-intersection coordinated control algorithm based on game theory and maximal flow,'' in \emph{Proc. IECON 39th Annu. Conf. IEEE Ind. Electron. Soc.}, Vienna, Nov. 2013, pp. 3258--3263, doi: {\color{blue}\href{http://dx.doi.org/10.1109/IECON.2013.6699650} {10.1109/IECON.2013.6699650}}.

\bibitem{Adkins2019game}
R.~P. Adkins, D.~M. Mount, and A.~A. Zhang, ``A game-theoretic approach for minimizing delays in autonomous intersections,'' in \emph{Traffic and Granular Flow '17}, S.~H. Hamdar, Ed.\hskip 1em plus 0.5em minus 0.4em\relax Cham: Springer International Publishing, 2019, pp. 131--139.

\bibitem{MaloTamayo2009}
A.~J. {Malo Tamayo}, A.~Poznyak, and I.~{\'{A}}. Villalobos, ``{Optimization and game at an urban intersection},'' \emph{IFAC Proc. Vol.}, no. 1983, pp. 526--530, 2009, doi: {\color{blue}\href{http://dx.doi.org/10.3182/20090902-3-US-2007.0030} {10.3182/20090902-3-US-2007.0030}}.

\bibitem{Guo2020}
J.~Guo and I.~Harmati, ``{Evaluating semi-cooperative Nash/Stackelberg Q-learning for traffic routes plan in a single intersection},'' \emph{Control Eng. Pract.}, vol. 102, no. June, p. 104525, 2020, doi: {\color{blue}\href{http://dx.doi.org/10.1016/j.conengprac.2020.104525} {10.1016/j.conengprac.2020.104525}}.

\bibitem{2015A}
M.~Mashayekhi and G.~F. List, ``A multiagent auction-based approach for modeling of signalized intersections,'' in \emph{Proc. IJCAI Workshops Synergies Between Multiagent Syst., Mach. Learn. Complex Syst.}, 2015.

\bibitem{guo2019optimization}
J.~Guo and I.~Harmati, ``Optimization of traffic signal control with different game theoretical strategies,'' in \emph{Proc. 2019 23rd Int. Conf. Syst. Theory, Control Comput.}, Sinaia, Romania, Oct. 2019, pp. 750--755, doi: {\color{blue}\href{http://dx.doi.org/10.1109/ICSTCC.2019.8885458} {10.1109/ICSTCC.2019.8885458}}.

\bibitem{Moya2015}
S.~Moya and J.~Escobar, ``{Stackelberg-Nash equilibrium in a traffic control problem at an intersection on a priority road},'' \emph{IMA J. Math. Control Inf.}, vol.~32, no.~1, pp. 161--194, 2015, doi: {\color{blue}\href{http://dx.doi.org/10.1093/imamci/dnt036} {10.1093/imamci/dnt036}}.

\bibitem{Guo2020many}
J.~Guo and I.~Harmati, ``{Comparison of game theoretical strategy and reinforcement learning in traffic light control},'' \emph{Periodica Polytechnica Transp. Eng.}, vol.~48, no.~4, pp. 313--319, 2020, doi: {\color{blue}\href{http://dx.doi.org/10.3311/PPTR.15923} {10.3311/PPTR.15923}}.

\bibitem{2014Game}
Y.~Narahari, \emph{Game theory and mechanism design}.\hskip 1em plus 0.5em minus 0.4em\relax World Scientific, 2014, vol.~4.

\bibitem{hayward1972near}
J.~C. Hayward, ``Near-miss determination through use of a scale of danger,'' \emph{Highway Res. Rec.}, no. 384, 1972.

\bibitem{zhang2012pedestrian}
Y.~Zhang, D.~Yao, T.~Z. Qiu, L.~Peng, and Y.~Zhang, ``Pedestrian safety analysis in mixed traffic conditions using video data,'' \emph{{IEEE} Trans. Intell. Transp. Syst.}, vol.~13, no.~4, pp. 1832--1844, 2012, doi: {\color{blue}\href{http://dx.doi.org/10.1109/TITS.2012.2210881} {10.1109/TITS.2012.2210881}}.

\bibitem{varhelyi1998drivers}
A.~Varhelyi, ``Drivers' speed behaviour at a zebra crossing: a case study,'' \emph{Accid. Anal. Prev.}, vol.~30, no.~6, pp. 731--743, 1998, doi: {\color{blue}\href{https://doi.org/10.1016/S0001-4575(98)00026-8} {10.1016/S0001-4575(98)00026-8}}.

\bibitem{abdelghaffar2019novel}
H.~M. Abdelghaffar and H.~A. Rakha, ``A novel decentralized game-theoretic adaptive traffic signal controller: Large-scale testing,'' \emph{Sensors}, vol.~19, no.~10, p. 2282, 2019, doi: {\color{blue}\href{http://dx.doi.org/10.3390/s19102282} {10.3390/s19102282}}.

\bibitem{hao2014cycle}
P.~Hao, X.~J. Ban, D.~Guo, and Q.~Ji, ``Cycle-by-cycle intersection queue length distribution estimation using sample travel times,'' \emph{Transp. Res. Part B, Methodol.}, vol.~68, pp. 185--204, 2014, doi: {\color{blue} \href{https://doi.org/10.1016/j.trb.2014.06.004} {10.1016/j.trb.2014.06.004}}.

\bibitem{ban2011real}
X.~J. Ban, P.~Hao, and Z.~Sun, ``Real time queue length estimation for signalized intersections using travel times from mobile sensors,'' \emph{Transp. Res. Part C, Emerg. Technol.}, vol.~19, no.~6, pp. 1133--1156, 2011, doi: {\color{blue} \href{https://doi.org/10.1016/j.trc.2011.01.002} {10.1016/j.trc.2011.01.002}}.

\bibitem{barth1996modal}
M.~Barth, F.~An, J.~Norbeck, and M.~Ross, ``Modal emissions modeling: A physical approach,'' \emph{Transp. Res. Rec.}, vol. 1520, no.~1, pp. 81--88, 1996, doi: {\color{blue}\href{https://doi.org/10.1177/0361198196152000110} {10.1177/0361198196152000110}}.

\bibitem{rakha2004development}
H.~Rakha, K.~Ahn, and A.~Trani, ``{Development of VT-Micro model for estimating hot stabilized light duty vehicle and truck emissions},'' \emph{Transp. Res. Part D, Transp. Environ.}, vol.~9, no.~1, pp. 49--74, 2004, doi: {\color{blue} \href{https://doi.org/10.1016/S1361-9209(03)00054-3} {10.1016/S1361-9209(03)00054-3}}.

\bibitem{wang2020movestar}
Z.~Wang, G.~Wu, and G.~Scora, ``Movestar: An open-source vehicle fuel and emission model based on usepa moves,'' \emph{arXiv preprint}, 2020, doi: {\color{blue}\href{http://dx. doi.org/10.48550/arXiv.2008.04986} {10.48550/arXiv.2008.04986}}.

\bibitem{barkley2019economics}
A.~Barkley, \emph{The economics of food and agricultural markets}.\hskip 1em plus 0.5em minus 0.4em\relax New Prairie Press, 2019.

\bibitem{Quincampoix2012}
M.~Quincampoix, \emph{Differential Games}.\hskip 1em plus 0.5em minus 0.4em\relax New York, NY: Springer New York, 2012, pp. 854--861, doi: {\color{blue} \href{https://doi.org/10.1007/978-1-4614-1800-9\_55} {10.1007/978-1-4614-1800-9\_55}}.

\bibitem{mkiramweni2019survey}
M.~E. Mkiramweni, C.~Yang, J.~Li, and W.~Zhang, ``A survey of game theory in unmanned aerial vehicles communications,'' \emph{IEEE Commun. Surveys Tuts.}, vol.~21, no.~4, pp. 3386--3416, 2019, doi: {\color{blue}\href{ //https://doi.org/10.1109/COMST.2019.2919613} {10.1109/COMST.2019.2919613}}.

\bibitem{driessen2013cooperative}
T.~S. Driessen, \emph{Cooperative games, solutions and applications}.\hskip 1em plus 0.5em minus 0.4em\relax Springer Science \& Business Media, 2013, vol.~3.

\bibitem{Wilson2011}
J.~Wilson, ``Cooperative games with transferable utility,'' in \emph{Wiley Encyclopedia of Operations Research and Management Science}.\hskip 1em plus 0.5em minus 0.4em\relax John Wiley \& Sons, Ltd, 2011, doi: {\color{blue} \href{http://doi.org/10.1002/9780470400531.eorms0199} {10.1002/9780470400531.eorms0199}}.

\bibitem{maschler2020game}
M.~Maschler, S.~Zamir, and E.~Solan, \emph{Game theory}.\hskip 1em plus 0.5em minus 0.4em\relax Cambridge University Press, 2020.

\bibitem{nash1950bargaining}
J.~F. Nash~Jr, ``The bargaining problem,'' \emph{Econometrica: J. Econometric Soc.}, pp. 155--162, 1950, doi: {\color{blue}\href{ //https://doi.org/10.2307/1907266} {10.2307/1907266}}.

\bibitem{milgrom2004putting}
P.~Milgrom and P.~R. Milgrom, \emph{Putting auction theory to work}.\hskip 1em plus 0.5em minus 0.4em\relax Cambridge University Press, 2004.

\bibitem{nash1996non}
J.~Nash~Jr, ``Non-cooperative games,'' in \emph{Essays on Game Theory}.\hskip 1em plus 0.5em minus 0.4em\relax Edward Elgar Publishing, 1996, pp. 22--33.

\bibitem{luo2022core}
C.~Luo, X.~Zhou, and B.~Lev, ``Core, shapley value, nucleolus and nash bargaining solution: A survey of recent developments and applications in operations management,'' \emph{Omega}, p. 102638, 2022, doi: {\color{blue}\href{ //https://doi.org/10.1016/j.omega.2022.102638} {10.1016/j.omega.2022.102638}}.

\bibitem{bowling2002multiagent}
M.~Bowling and M.~Veloso, ``Multiagent learning using a variable learning rate,'' \emph{Artif. Intell.}, vol. 136, no.~2, pp. 215--250, 2002, doi: {\color{blue}\href{ https://doi.org/10.1016/S0004-3702(02)00121-2} {10.1016/S0004-3702(02)00121-2}}.

\bibitem{balaji2010introduction}
P.~G. Balaji and D.~Srinivasan, ``An introduction to multi-agent systems,'' in \emph{Innovations in multi-agent systems and applications-1}.\hskip 1em plus 0.5em minus 0.4em\relax Springer, 2010, pp. 1--27.

\bibitem{nowe2012game}
A.~Now{\'e}, P.~Vrancx, and Y.-M.~D. Hauwere, ``Game theory and multi-agent reinforcement learning,'' in \emph{Reinforcement Learning}.\hskip 1em plus 0.5em minus 0.4em\relax Springer, 2012, pp. 441--470.

\bibitem{el2013multiagent}
S.~El-Tantawy, B.~Abdulhai, and H.~Abdelgawad, ``Multiagent reinforcement learning for integrated network of adaptive traffic signal controllers (marlin-atsc): methodology and large-scale application on downtown toronto,'' \emph{{IEEE} Trans. Intell. Transp. Syst.}, vol.~14, no.~3, pp. 1140--1150, 2013, doi: {\color{blue}\href{ https://doi.org/10.1109/TITS.2013.2255286} {10.1109/TITS.2013.2255286}}.

\bibitem{yang2020overview}
Y.~Yang and J.~Wang, ``An overview of multi-agent reinforcement learning from game theoretical perspective,'' \emph{arXiv preprint}, 2020, doi: {\color{blue} \href{ https://doi.org/10.48550/arXiv.2011.00583} {10.48550/arXiv.2011.00583}}.

\bibitem{busoniu2008comprehensive}
L.~Busoniu, R.~Babuska, and B.~De~Schutter, ``A comprehensive survey of multiagent reinforcement learning,'' \emph{IEEE Trans. Syst., Man, Cybern. C, Appl. Rev.}, vol.~38, no.~2, pp. 156--172, 2008, doi: {\color{blue} \href{ https://doi.org/10.1109/TSMCC.2007.913919} {10.1109/TSMCC.2007.913919}}.

\bibitem{vinyals2017starcraft}
O.~Vinyals, T.~Ewalds, S.~Bartunov, P.~Georgiev, A.~S. Vezhnevets, M.~Yeo, A.~Makhzani, H.~K{\"u}ttler, J.~Agapiou, J.~Schrittwieser \emph{et~al.}, ``Starcraft ii: A new challenge for reinforcement learning,'' \emph{arXiv preprint}, 2017, doi: {\color{blue}\href{ https://doi.org/10.48550/arXiv.1708.04782} {10.48550/arXiv.1708.04782}}.

\bibitem{terry2021pettingzoo}
J.~Terry, B.~Black, N.~Grammel, M.~Jayakumar, A.~Hari, R.~Sullivan, L.~Santos, R.~Perez, C.~Horsch, C.~Dieffendahl \emph{et~al.}, ``Pettingzoo: Gym for multi-agent reinforcement learning,'' \emph{arXiv e-prints}, 2020, doi: {\color{blue}\href{ https://doi.org/10.48550/arXiv.2009.14471} {10.48550/arXiv.2009.14471}}.

\bibitem{zhang2021multi}
K.~Zhang, Z.~Yang, and T.~Ba{\c{s}}ar, ``Multi-agent reinforcement learning: A selective overview of theories and algorithms,'' in \emph{Handbook of Reinforcement Learning and Control}.\hskip 1em plus 0.5em minus 0.4em\relax Springer, 2021, pp. 321--384.

\bibitem{bazzan2009opportunities}
A.~L. Bazzan, ``Opportunities for multiagent systems and multiagent reinforcement learning in traffic control,'' \emph{Auton. Agents Multi-Agent Syst.}, vol.~18, no.~3, pp. 342--375, 2009, doi: {\color{blue}\href{https://doi.org/10.1007/s10458-008-9062-9} {10.1007/s10458-008-9062-9}}.

\bibitem{shapley1953stochastic}
L.~S. Shapley, ``Stochastic games,'' \emph{Proc. Nat. Acad. Sci.}, vol.~39, no.~10, pp. 1095--1100, 1953, doi: {\color{blue}\href{ https://doi.org/10.1073/pnas.39.10.1095} {10.1073/pnas.39.10.1095}}.

\bibitem{dresner2006multiagent}
K.~Dresner and P.~Stone, ``Multiagent traffic management: Opportunities for multiagent learning,'' in \emph{Proc. Learn. Adaption Multiagent Syst.}\hskip 1em plus 0.5em minus 0.4em\relax Springer, 2006, pp. 129--138, doi: {\color{blue}\href{http://dx.doi.org/10.1007/11691839} {10.1007/11691839}}.

\bibitem{koonce2008traffic}
P.~Koonce and L.~Rodegerdts, ``Traffic signal timing manual.'' United States. Federal Highway Administration, Tech. Rep., 2008.

\bibitem{varaiya2013max}
P.~Varaiya, ``The max-pressure controller for arbitrary networks of signalized intersections,'' in \emph{Advances in Dynamic Network Modeling in Complex Transportation Systems}.\hskip 1em plus 0.5em minus 0.4em\relax Springer, 2013, pp. 27--66.

\bibitem{roess2011traffic}
R.~P. Roess, E.~S. Prassas, and W.~R. McShane, \emph{Traffic engineering}.\hskip 1em plus 0.5em minus 0.4em\relax Prentice Hall, 2011.

\bibitem{wu2019dcl}
Y.~Wu, H.~Chen, and F.~Zhu, ``Dcl-aim: Decentralized coordination learning of autonomous intersection management for connected and automated vehicles,'' \emph{Transp. Res. Part C, Emerg. Technol.}, vol. 103, pp. 246--260, 2019, doi: {\color{blue}\href{ https://doi.org/10.1016/j.trc.2019.04.012} {10.1016/j.trc.2019.04.012}}.

\bibitem{robertson1991optimizing}
D.~I. Robertson and R.~D. Bretherton, ``Optimizing networks of traffic signals in real time-the scoot method,'' \emph{{IEEE} Trans. Veh. Technol.}, vol.~40, no.~1, pp. 11--15, 1991, doi: {\color{blue}\href{ https://doi.org/10.1109/25.69966} {10.1109/25.69966}}.

\bibitem{cools2013self}
S.-B. Cools, C.~Gershenson, and B.~D'Hooghe, \emph{Self-Organizing Traffic Lights: A Realistic Simulation}.\hskip 1em plus 0.5em minus 0.4em\relax London: Springer London, 2013, pp. 45--55, doi: {\color{blue}\href{ https://doi.org/10.1007/978-1-4471-5113-5\_3} {10.1007/978-1-4471-5113-5\_3}}.

\bibitem{wei2019presslight}
H.~Wei, C.~Chen, G.~Zheng, K.~Wu, V.~Gayah, K.~Xu, and Z.~Li, ``Presslight: Learning max pressure control to coordinate traffic signals in arterial network,'' in \emph{Proc. 25th ACM SIGKDD Int. Conf. Knowl. Discovery Data Mining}, 2019, pp. 1290--1298, doi: {\color{blue}\href{ https://https://doi.org/10.1145/3292500.3330949} {10.1145/3292500.3330949}}.

\bibitem{wei2019colight}
H.~Wei, N.~Xu, H.~Zhang, G.~Zheng, X.~Zang, C.~Chen, W.~Zhang, Y.~Zhu, K.~Xu, and Z.~Li, ``Colight: Learning network-level cooperation for traffic signal control,'' in \emph{Proc. 28th ACM Int. Conf. Inf. Knowl. Manage.}, 2019, pp. 1913--1922, doi: {\color{blue}\href{ https://https://doi.org/10.1145/3357384.3357902} {10.1145/3357384.3357902}}.

\bibitem{chu2019multi}
T.~Chu, J.~Wang, L.~Codec{\`a}, and Z.~Li, ``Multi-agent deep reinforcement learning for large-scale traffic signal control,'' \emph{{IEEE} Trans. Intell. Transp. Syst.}, vol.~21, no.~3, pp. 1086--1095, 2019, doi: {\color{blue} \href{ https://doi.org/10.1109/TITS.2019.2901791} {10.1109/TITS.2019.2901791}}.

\bibitem{zang2020metalight}
X.~Zang, H.~Yao, G.~Zheng, N.~Xu, K.~Xu, and Z.~Li, ``Metalight: Value-based meta-reinforcement learning for traffic signal control,'' in \emph{Proc. AAAI Conf. Artif. Intell.}, vol.~34, no.~01, 2020, pp. 1153--1160, doi: {\color{blue}\href{ https://https://doi.org/10.1609/aaai.v34i01.5467} {10.1609/aaai.v34i01.5467}}.

\bibitem{devailly2021ig}
F.-X. Devailly, D.~Larocque, and L.~Charlin, ``Ig-rl: Inductive graph reinforcement learning for massive-scale traffic signal control,'' \emph{{IEEE} Trans. Intell. Transp. Syst.}, vol.~23, no.~7, pp. 7496--7507, 2021, doi: {\color{blue}\href{ https://https://doi.org/10.1109/TITS.2021.3070835} {10.1109/TITS.2021.3070835}}.

\bibitem{chen2020toward}
C.~Chen, H.~Wei, N.~Xu, G.~Zheng, M.~Yang, Y.~Xiong, K.~Xu, and Z.~Li, ``Toward a thousand lights: Decentralized deep reinforcement learning for large-scale traffic signal control,'' in \emph{Proc. AAAI Conf. Artif. Intell.}, vol.~34, no.~04, 2020, pp. 3414--3421, doi: {\color{blue}\href{ https://https://doi.org/10.1609/aaai.v34i04.5744} {10.1609/aaai.v34i04.5744}}.

\bibitem{oroojlooy2020attendlight}
\BIBentryALTinterwordspacing
A.~Oroojlooy, M.~Nazari, D.~Hajinezhad, and J.~Silva, ``Attendlight: Universal attention-based reinforcement learning model for traffic signal control,'' in \emph{Proc. Adv. Neural Inf. Process. Syst.}, H.~Larochelle, M.~Ranzato, R.~Hadsell, M.~Balcan, and H.~Lin, Eds., vol.~33.\hskip 1em plus 0.5em minus 0.4em\relax Curran Associates, Inc., 2020, pp. 4079--4090. [Online]. Available: \url{https://proceedings.neurips.cc/paper_files/paper/2020/file/29e48b79ae6fc68e9b6480b677453586-Paper.pdf}
\BIBentrySTDinterwordspacing

\bibitem{lopez2018microscopic}
P.~A. Lopez, M.~Behrisch, L.~Bieker-Walz, J.~Erdmann, Y.-P. Fl{\"o}tter{\"o}d, R.~Hilbrich, L.~L{\"u}cken, J.~Rummel, P.~Wagner, and E.~Wie{\ss}ner, ``Microscopic traffic simulation using sumo,'' in \emph{Proc. 21st Int. Conf. Intell. Transp. Syst. (ITSC)}, Maui, HI, USA, Nov. 2018, pp. 2575--2582, doi: {\color{blue}\href{ https://https://doi.org/10.1109/ITSC.2018.8569938} {10.1109/ITSC.2018.8569938}}.

\bibitem{cameron1996paramics}
G.~D. Cameron and G.~I. Duncan, ``Paramics—parallel microscopic simulation of road traffic,'' \emph{J. Supercomput.}, vol.~10, pp. 25--53, 1996, doi: {\color{blue}\href{ https://https://doi.org/10.1007/BF00128098} {10.1007/BF00128098}}.

\bibitem{casas2010traffic}
J.~Casas, J.~L. Ferrer, D.~Garcia, J.~Perarnau, and A.~Torday, \emph{Traffic Simulation with Aimsun}.\hskip 1em plus 0.5em minus 0.4em\relax New York, NY: Springer New York, 2010, pp. 173--232, doi: {\color{blue}\href{ https://https://10.1007/978-1-4419-6142-6\_5} {10.1007/978-1-4419-6142-6\_5}}.

\bibitem{fellendorf2010microscopic}
M.~Fellendorf and P.~Vortisch, \emph{Microscopic Traffic Flow Simulator VISSIM}.\hskip 1em plus 0.5em minus 0.4em\relax New York, NY: Springer New York, 2010, pp. 63--93, doi: {\color{blue}\href{ https://https://doi.org/10.1007/978-1-4419-6142-6\_2} {10.1007/978-1-4419-6142-6\_2}}.

\bibitem{zhang2019cityflow}
H.~Zhang, S.~Feng, C.~Liu, Y.~Ding, Y.~Zhu, Z.~Zhou, W.~Zhang, Y.~Yu, H.~Jin, and Z.~Li, ``Cityflow: A multi-agent reinforcement learning environment for large scale city traffic scenario,'' in \emph{Proc. World Wide Web Conf.}, 2019, pp. 3620--3624, doi: {\color{blue}\href{ https://https://doi.org/10.1145/3308558.3314139} {10.1145/3308558.3314139}}.

\bibitem{wu2021flow}
C.~Wu, A.~R. Kreidieh, K.~Parvate, E.~Vinitsky, and A.~M. Bayen, ``Flow: A modular learning framework for mixed autonomy traffic,'' \emph{{IEEE} Trans. Robot.}, vol.~38, no.~2, pp. 1270--1286, 2021, doi: {\color{blue}\href{ https://https://doi.org/10.1109/TRO.2021.3087314} {10.1109/TRO.2021.3087314}}.

\bibitem{garg2019traffic3d}
D.~Garg, M.~Chli, and G.~Vogiatzis, ``Traffic3d: A new traffic simulation paradigm,'' in \emph{Proc. Int. Jt. Conf. Auton. Agents Multiagent Syst.}, 2019, pp. 2354--2356, doi: {\color{blue}\href{ //https://doi.org/10.5555/3306127.3332110} {10.5555/3306127.3332110}}.

\bibitem{zhou2020smarts}
M.~Zhou, J.~Luo, J.~Villella, Y.~Yang, D.~Rusu, J.~Miao, W.~Zhang, M.~Alban, I.~Fadakar, Z.~Chen \emph{et~al.}, ``Smarts: Scalable multi-agent reinforcement learning training school for autonomous driving,'' \emph{arXiv preprint}, 2020, doi: {\color{blue}\href{ https://doi.org/10.48550/arXiv.2010.09776 } { 10.48550/arXiv.2010.09776 }}.

\bibitem{highway-env}
E.~Leurent, ``An environment for autonomous driving decision-making,'' \url{https://github.com/eleurent/highway-env}, 2018.

\bibitem{Movesim2010}
M.~Treiber and A.~Kesting, ``An open-source microscopic traffic simulator,'' \emph{{IEEE} Intell. Transp. Syst. Mag.}, vol.~2, no.~3, pp. 6--13, 2010, doi: {\color{blue}\href{ //https://doi.org/10.1109/MITS.2010.939208} {10.1109/MITS.2010.939208}}.

\bibitem{dosovitskiy2017carla}
\BIBentryALTinterwordspacing
A.~Dosovitskiy, G.~Ros, F.~Codevilla, A.~Lopez, and V.~Koltun, ``Carla: An open urban driving simulator,'' in \emph{Proc. Conf. Robot Learn.}\hskip 1em plus 0.5em minus 0.4em\relax PMLR, 2017, pp. 1--16. [Online]. Available: \url{https://proceedings.mlr.press/v78/dosovitskiy17a.html}
\BIBentrySTDinterwordspacing

\bibitem{bowling2000analysis}
M.~Bowling and M.~Veloso, ``An analysis of stochastic game theory for multiagent reinforcement learning,'' Carnegie Mellon Univ., Pittsburgh, PA, USA, Tech. Rep. No. CMU-CS00-165, 2000.

\bibitem{martinez2010emergency}
F.~J. Martinez, C.-K. Toh, J.-C. Cano, C.~T. Calafate, and P.~Manzoni, ``Emergency services in future intelligent transportation systems based on vehicular communication networks,'' \emph{{IEEE} Intell. Transp. Syst. Mag.}, vol.~2, no.~2, pp. 6--20, 2010, doi: {\color{blue} \href{https://doi.org/10.1109/MITS.2010.938166} {10.1109/MITS.2010.938166}}.

\end{thebibliography}

\section{Biography Section}

\begin{IEEEbiography}[{\includegraphics[width=1in,height=1.25in,clip,keepaspectratio]{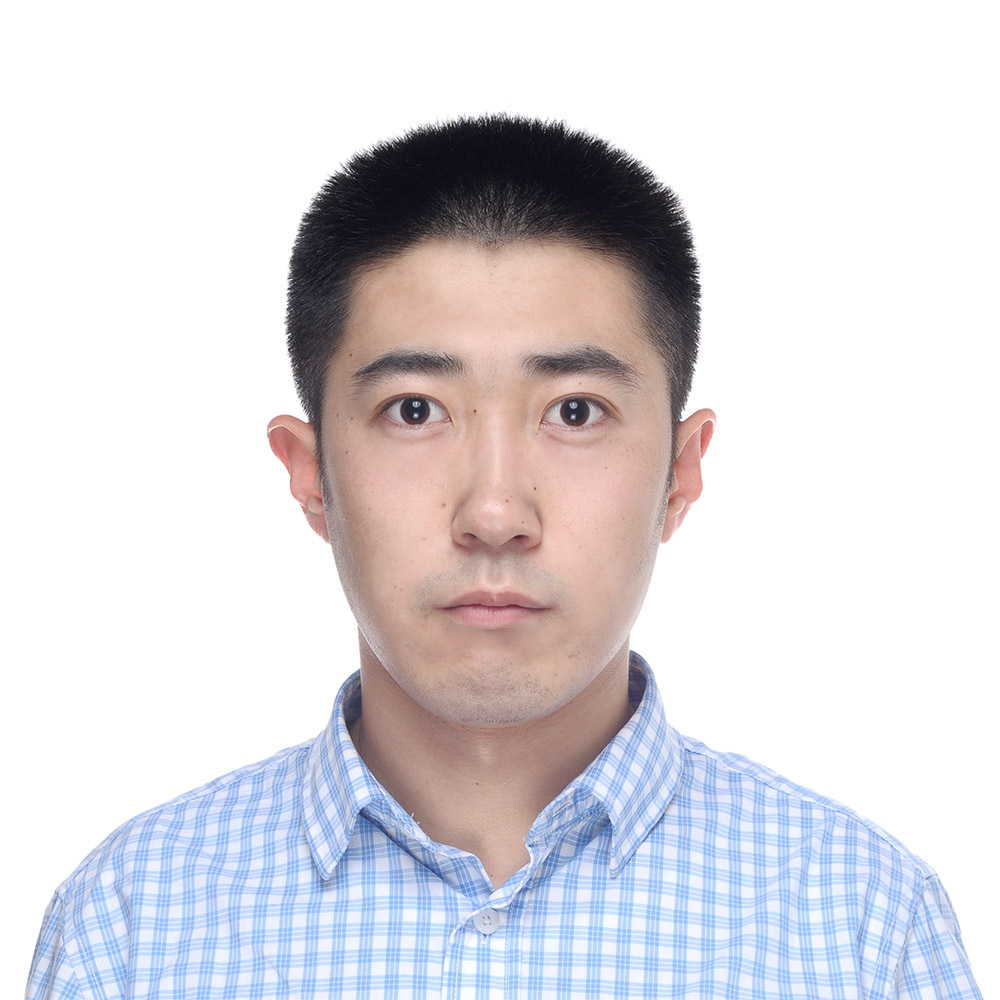}}]{Ziye Qin} (S’23) is currently pursuing the Ph.D. degree at the School of Transportation and Logistics, Southwest Jiaotong University, Chengdu, China. He is also a visiting student with the College of Engineering,
Center for Environmental Research and Technology, University of California at Riverside, Riverside, CA, USA. His research interests include game theory and mechanism design, motion planning and cooperative driving automation.
\end{IEEEbiography}

\begin{IEEEbiography}[{\includegraphics[width=1in,height=1.25in,clip,keepaspectratio]{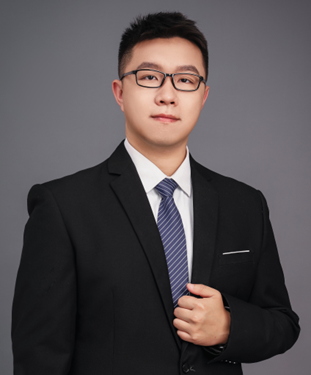}}]{Ang Ji} (M’23) received the Ph.D. degree from the School of Civil Engineering, The University of Sydney, Australia, in 2022. He now serves as an Assistant Professor at the School of Transportation and Logistics, Southwest Jiaotong University, Chengdu, China. His research interests include connected automated vehicles, deep/reinforcement learning, microscopic traffic modeling, and game-theoretic applications.
\end{IEEEbiography}

\begin{IEEEbiography}[{\includegraphics[width=1in,height=1.25in,clip,keepaspectratio]{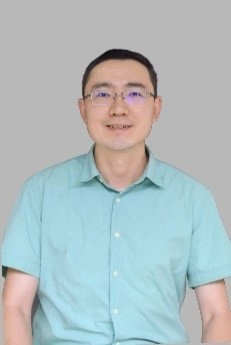}}]{Zhanbo Sun} received his Ph.D. degree from Rensselaer Polytechnic Institute in 2014 and the B.E. degree from Tsinghua University in 2009. He is now a Professor at the School of Transportation and Logistics in Southwest Jiaotong University. His research focuses on intelligent connected vehicles, green transportation, and rail transit. He has won the Jeme Tianyou Railway Science and Technology Award and the Best Paper Award of the International Federation of Automatic Control (IFAC TA 2019).
\end{IEEEbiography}

\begin{IEEEbiography}[{\includegraphics[width=1in,height=1.25in,clip,keepaspectratio]{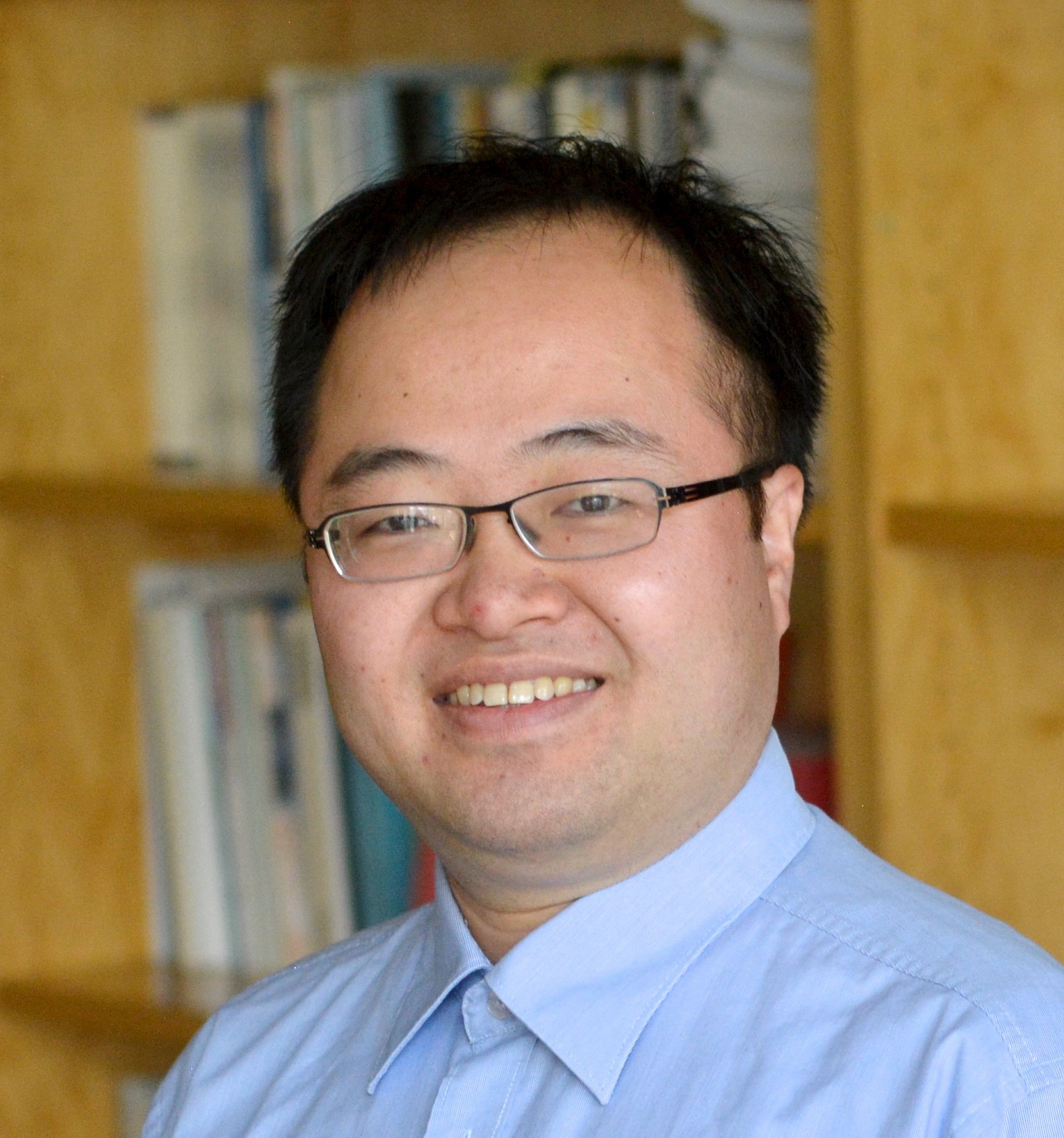}}]{Guoyuan Wu} (M’09-SM’15) received his Ph.D. degree in mechanical engineering from the University of California, Berkeley in 2010. Currently, he holds a Full Researcher and Adjunct Professor position at Bourns College of Engineering – Center for Environmental Research \& Technology (CE–CERT) and Department of Electrical \& Computer Engineering in the University of California at Riverside. His research interest has been focused on the development and evaluation of sustainable and intelligent transportation system (SITS) technologies, including connected and automated transportation systems (CATS), shared mobility, transportation electrification, optimization and control of vehicles, traffic simulation, and energy/emissions measurement and modeling. Dr. Wu serves as an Associate Editor for a few prestigious journals. He is a member of National Academy of Inventors (NAI) and a member of a few Standing Committees of the Transportation Research Board (TRB). He is the recipient of 2020 Vincent Bendix Automotive Electronics Engineering Award, 2021 Arch T. Colwell Merit Award, and 2017-2020 IEEE-ITSM Outstanding Survey Paper Award. 
\end{IEEEbiography}

\begin{IEEEbiography}[{\includegraphics[width=1in,height=1.25in,clip,keepaspectratio]{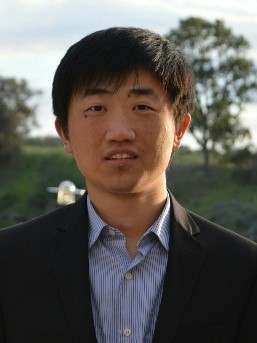}}]{Peng Hao} (M’16) received Ph.D. degree in transportation engineering from Rensselaer Polytechnic Institute in 2013. He is currently an Assistant Research Engineer with the Transportation Systems Research Group, Center for Environmental Research and Technology, Bourns College of Engineering, University of California, Riverside, USA. His research interests include connected vehicles, eco-approach and departure, sensor-aided modeling, shared mobility and traffic operations.
\end{IEEEbiography}

\begin{IEEEbiography}[{\includegraphics[width=1in,height=1.25in,clip,keepaspectratio]{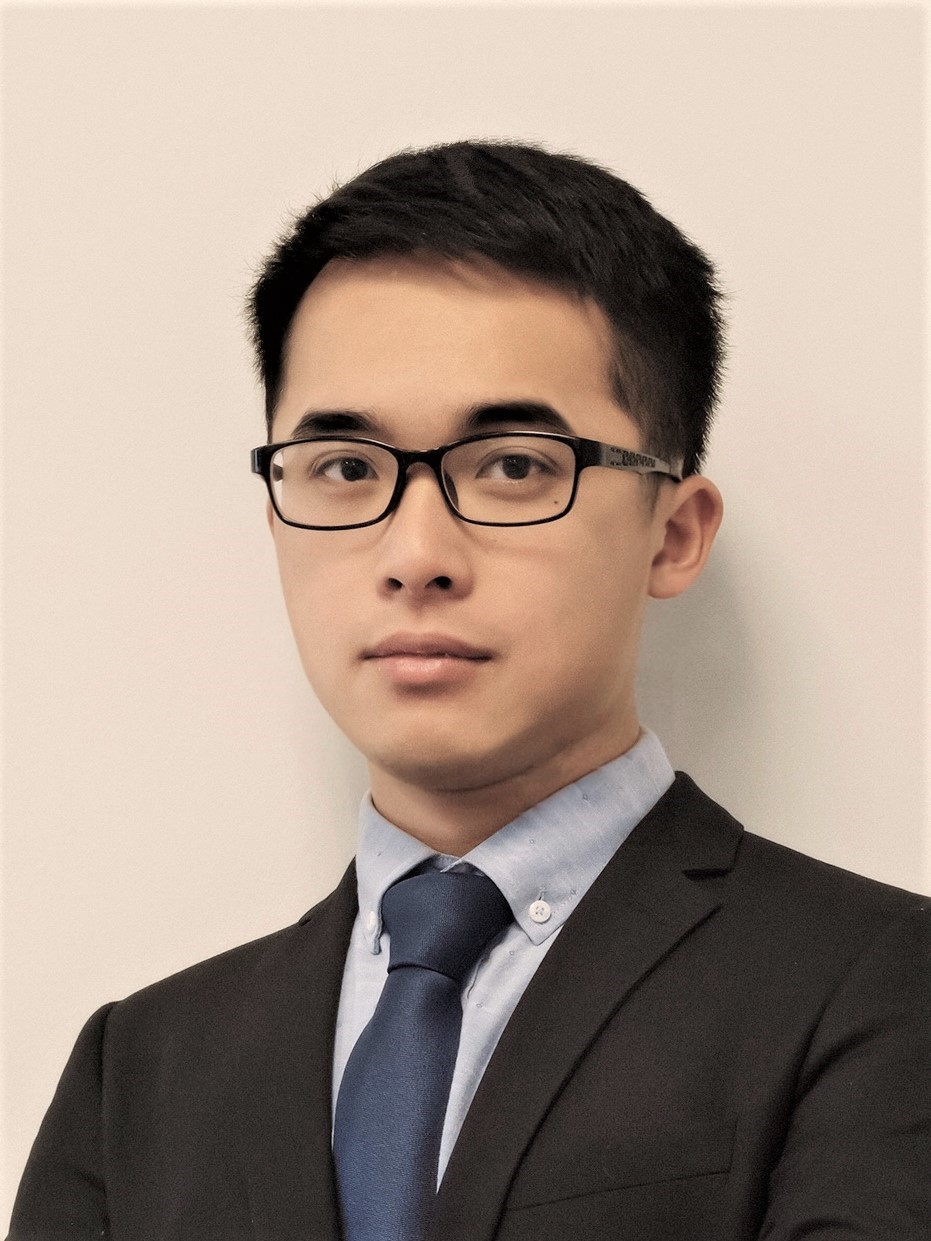}}]{Xishun Liao} (S’19) received the Ph.D. degree in electrical and computer engineering from University of California at Riverside in 2023, the M.Eng. degree in mechanical engineering from University of Maryland at College Park in 2018, and the B.E. degree in mechanical engineering and automation from Beijing University of Posts and Telecommunications in 2016. He is currently a postdoctoral scholar at the University of California at Los Angeles. His research focuses on motion planning and control, driver behavior, intelligent transportation system technologies.
\end{IEEEbiography}

\vspace{11pt}

\vfill

\end{document}